\documentstyle{article}
\newcommand{\bildh}[3]{{\picplace{#2}}\label{#1}}

\newcommand{\thetitle}{}
\newcommand{\thesubtitle}{}
\newcommand{\thedate}{}
\newcommand{\theauthor}{}
\newcommand{\theinstitute}{}

\renewcommand{\title}[1]{\renewcommand{\thetitle}{#1}}
\newcommand{\subtitle}[1]{\renewcommand{\thesubtitle}{\par #1}}
\renewcommand{\date}[1]{\renewcommand{\thedate}{#1}}
\renewcommand{\author}[1]{\renewcommand{\theauthor}{#1}}
\newcommand{\institute}[1]{\renewcommand{\theinstitute}{#1}}

\newcommand{\inst}[1]{$^{#1}$\renewcommand{\and}{, }}
\newcommand{\instn}[1]{$^{#1}$\renewcommand{\and}{\\}}
\newcommand{\thesaurus}[1]{}
\newcommand{\offprints}[1]{\renewcommand{\thefootnote}{}
\footnotetext{{\it Send offprints
requests to:} #1}\renewcommand{\thefootnote}{\arabic{footnote}}
}

\renewcommand{\maketitle}{
\thispagestyle{empty}
{\parindent0cm
\begin{center}
\vskip 2em
{\LARGE \thetitle \par {\Large \thesubtitle}}\par
\vskip 2em {\large \theauthor}\par
\vskip 1em {\theinstitute}\par
\vskip 2em {\thedate}
\end{center}
}}

\newcommand{\la}{\le}
\newcommand{\ga}{\ge}
\newcommand{\keywords}{{\bf Keywords:} }
\newcommand{\arcsec}{\mbox{''}}
\newcommand{\arcmin}{\mbox{'}}
\newcommand{\degr}{\mbox{$^\circ$}}
\newcommand{\picplace}[1]{\frame{\centerline{(will be inserted later)}}}

\renewenvironment{thebibliography}[1]{{\section*{References}}
\parindent0cm}{\par\vskip1.5em}
\renewcommand{\bibitem}[7]{\par}

\begin{document}

\thesaurus{03(11.01.2; 11.10.1; 11.14.1; 02.01.2)}
\title{The jet-disk symbiosis}
\subtitle{II. Interpreting the Radio/UV correlations in quasars}
\author{Heino Falcke\inst{1}\and Matthew A. Malkan\inst{2}\and Peter L.
Biermann\inst{1}}
\offprints{HFALCKE@mpifr-bonn.mpg.de}
\institute{Max-Planck Institut f\"ur Radioastronomie, Auf dem H\"ugel 69,
D-53121 Bonn, Germany\instn{1} \and Department of Astronomy, UCLA, Los Angeles,
CA 90024-1562, USA\instn{2}}
\date{Astronomy \& Astrophysics, in press [astro-ph/9411100]}
\maketitle
\markboth{Falcke, Malkan, Biermann: The jet-disk symbiosis II. Radio/UV
correlations in quasars}{Falcke, Malkan, Biermann: The jet-disk symbiosis II.
Radio/UV correlations in quasars}
\begin{abstract}
We investigate the correlation between the accretion disk (UV)
luminosity and the radio core emission of a quasar sample, containing
all PG quasars, also deriving empirical conversion factors from
emission line luminosities to disk luminosities. This method allows us
to investigate the radio properties of AGN on the absolute scale set
by the accretion power.  In a radio vs. $L_{\rm disk}$ plot we find
the quasars to be separated into four classes: core dominated quasars
(CDQ), lobe dominated quasars (LDQ), radio-intermediate quasars (RIQ)
and radio weak quasars. In general the radio core emission scales with
the disk luminosity, especially in the radio weak quasars.  This shows
that radio and UV emission have a common energy source and that the
difference between radio loud and radio weak is established already on
the parsec scale.  We investigate the possibility that radio jets are
responsible for the radio core emission in radio loud and radio weak
quasars. Comparing our data with a simple jet emission model that
takes the limits imposed by energy and mass conservation in a coupled
jet-disk system into account, we find that radio loud jets carry a
total power $Q_{\rm jet}$ that is at least 1/3 of the observed disk
luminosity $L_{\rm disk}$. The strong radio core emission in radio
loud quasars relative to $L_{\rm disk}$ is difficult to explain by
normal acceleration of thermal electrons into a non-thermal powerlaw
distribution. One rather is forced to postulate an efficient process
producing a large number of pairs and/or injecting electrons with a
distribution with low-energy cut-off around 50 MeV -- secondary pair
production in hadronic cascades could be such a process. Width and
shape of the UV-radio correlation of LDQ and CDQ limit the parameter
range for the bulk Lorentz factor of the jet to a narrow region
($3\la\gamma_{\rm j}\la10$)). The radio emission of radio weak quasars
can be explained with exactly the same parameters for a powerful
relativistic jet if secondary pair production, as suggested for radio
loud jets, is inhibited. There is evidence that RIQ are the
relativistically boosted population of radio weak quasar jets.  Our
diagram provides also a test for the unification of FR\,II radio
galaxies with quasars as we can estimate the hidden disk luminosity of
FR II galaxies and find that this is consistent with FRII being
misdirected quasars.

\keywords{galaxies: active -- galaxies: jets -- galaxies: nuclei -- accretion
disks}
\end{abstract}
\section{Introduction}
The standard picture for the active galactic nuclei (AGN) of galaxies
classified as quasars still is an accreting supermassive black hole
surrounded by an accretion disk. The release of energy due to the
infall in the gravitational potential is thought to be responsible for
the enormous energy output seen in the optical to ultraviolet (UV)
which is able to dominate the total emission of the underlying galaxy.
While for individual sources the spectral energy distribution (SED)
may be dominated by other contributions, e.g. gamma-rays, the UV
emission in general is assumed to be the major channel for the energy
release of quasars. High gamma-ray fluxes occur only in a few sources
and are obviously boosted. However, recently it was pointed out that
relativistic jets in radio loud quasars may be another important
channel for expelling high amounts of energy from the central engine
(Rawlings \& Saunders 1991, hereafter RS91; Falcke, Mannheim, Biermann
1993b, hereafter FMB; Falcke \& Biermann 1994b, hereafter {\em Paper
I}; Celotti \& Fabian 1993). Although the radiative contribution of
jets to the SED in the radio and gamma regime is small, their energy
flow in bulk relativistic motion, magnetic energy and relativistic
particles may be a substantial fraction of total energy released from
the central engine. Such a powerful jet suggests that this kind of
outflow is directly coupled to the accretion process in the very
vicinity of the black hole -- where most of the energy is released --
and will have a strong impact on the disk itself.  Therefore one can
not treat disk and jet as two isolated parts but must treat them as a
coupled sytem.

In previous papers ({\em Paper I}, FMB) we studied the standard theory
for the radio emission of compact radio cores produced by radio jets
in quasars considering especially the link between the jet and an
accretion disk as the basic source for energy and matter.  This is
simply the Blandford \& K\"onigl (1979) theory plus energy and mass
conservation in the jet-disk system.

A necessary consequence of this jet-disk link is the scaling of the
emission from the compact radio core in galactic nuclei with the disk
luminosity radiated in the UV where the scaling is not necessarily
linear.  Another conclusion was that for a given accretion rate, the
core emission of radio loud jets is already close to the maximum
efficiency, and brighter emission is only possible by relativistic
boosting.  Moreover Falcke \& Biermann (1994b) introduced the concept
of a jet-disk symbiosis and started with the basic Ansatz that jets
are necessary partners of accretion disks around compact objects, and
not fortuitous by-products of disks. In this context the question was
raised whether the radio emission of radio-weak quasars could be
explained by the same kind of relativistic jets as well but being less
efficient in accelerating electrons. The predictions are that UV and
radio core emission should be strongly correlated in radio weak
quasars as well and that boosted radio weak jets with enhanced
variable radio emission are found. Obviously depends the dichotomy of
the radio emission on the host galaxy (radio loud: elliptical, radio
weak: spiral) and it is hard to understand why the basic parameters of
jet flows should differ as a function of the host galaxy when the
production happens very close to the black hole where the effects of
the host galaxy should be negligible.

In this paper we now want to test some of these ideas and see if they
contradict available observational data.  If there is indeed a
fundamental process linking accretion process and jet production, it
should be insensitive to the environment and the host galaxy except
for the bimodality between radio loud and radio weak quasars.  Further
out the environmental impact of the host galaxy will inevitably change
the appearance of the jet and disk and modify the emission region.
Therefore one has to select carefully the observational data for each
object, and must make sure that the observed radiation reflects the
physical situation of the center and not of the outer region. To
achieve this we can follow two extreme strategies, one is to have very
high spatial resolution of the observation e.g. by looking at nearby
galactic nuclei. The other extreme would be to look at very luminous
objects, where the central engine outshines all other sources in a
galaxy like in quasars.  Here we will follow the second approach and
leave the nearby nuclei -- with the exception of our Galactic Center,
which is discussed in Falcke et al.  (1993a\&b), Falcke \& Biermann
(1994a\&c) and Falcke (1994a\&b), to future work. All other galactic
nuclei, e.g. Seyfert and radio galaxies, are very difficult to use as
test candidates as it requires an elaborated scheme to separate the
contributions from different parts of the galaxy (e.g.  starburst and
AGN).

We build on the work by Miller et al. (1993a, MSR) who investigated
the optically selected PG bright quasar sample and found correlations
between radio and optical line emission luminosities of radio weak
quasars.  We now extend this analysis by enlarging the sample with
more sources and comparing it with FR\,II galxies in light of current
unification theories. Moreover we try to determine the disk
luminosities of the quasars in our sample as accurately as possible
from existing data to compare disk and radio power in absolute terms.
This allows us to constrain strongly the parameter range of radio
jets and also give a solid classification for the radio properties of
quasars.

In Sec. 2 of this paper we introduce our sample and present our method
to estimate the disk luminosity of the individual quasars and
calibrate different other indicators to obtain empirical conversion
factors, e.g. from line luminosities to disk luminosities. Section 3
describes the available radio data and Sec. 4 the UV-radio diagram and
its implication for the different classes of quasars. In Sec. 5 we
discuss the constraints this diagram imposes on jet parameters like
Lorentz factor and total jet power. Section 5 discusses the
implication of our estimated parameters for the unification of quasars
and FR\,II radio galaxies and provides a simple method to estimate the
hidden disk luminosity in radio galaxies.

For consistency with some of the observational papers we use $q_0=.5$
and $H_0=50$ km/sec/Mpc as cosmological parameters.

\section{The disk luminosity}
\subsection{The quasar sample}
Most quasars show a fairly homogenous SED
with two important features --
an infrared (IR) and an UV-bump (Edelson and Malkan 1986;
Sanders et al. 1989). While the IR seems
to be due to thermal emission from dust, the dominant UV bump probably
reflects the spectrum of an accretion disk around a supermassive black
hole (Malkan \& Sargent 1982, Malkan 1983). We now estimate
this UV-bump luminosity -- which hereafter we will also refer to as
`disk luminosity' ($L_{\rm disk}$) -- as accurately as possible.

We take the observationally best studied sample of quasars namely the
PG sample of bright quasars (Schmidt \& Green 1983) with redshift
$z<0.5$ with a multitude of published observational data. Boroson \&
Green (1992, hereafter BG92) presented a thorough study of the
emission line properties of this sample while Sun \& Malkan (1989,
hereafter SM89) determined the disk luminosity from fitting disk
spectra to optical and UV data for a sample containg many of the PG
quasars.  Kellermann et al. (1989, hereafter KSSSG) and Miller et al.
(1993, hereafter MSR) published radio data obtained with the VLA. MSR
also tried to classify the morphological structure of the radio maps
and studied correlation between radio fluxes and emission lines.

The low redshift PG sample ($z<0.5$) we will mainly base our arguments
on is given in Table 1 and contains 87 sources (\#1-\#87). We did not
include the recent addition PG 1001+054 (not included in BG92) or PG
0119+229, since it shows no broad lines.  In addition we will consider
an extended sample which cannot be used for statistical purposes, as
it is not complete and has no clear defined selection criterion.
Sources \#88-\#113 in Tab. 1 are the PG quasars with $z>0.5$ which
were measured by KSSSG as well but not discussed by MRS.  It is known
that there are some selection effects for the PG sample with $z>0.5$
(Wampler \& Ponz 1985). Beyond the whole PG sample we have the sources
\#113-\#131 which (with the exception of 1821+643) were included in
the sample of SM89 and had disk fits available.

Besides quasars we will discuss briefly a sample of Fanaroff-Riley
type II (FRII) radio galaxies with $z<0.5$ which are selected from the
3C-based sample of Laing et al. (1983) and were investigated further
by Rawlings et al. (1989) and RS91. It contains 39 FR\,II galaxies and
is listed in Table 2 (\#132-\#170). The Laing et al.  (1983) sample is
96\% complete to a flux density limit of 10 Jy at 178 Mhz for sources
smaller than 10\arcmin and can be considered as fairly unbiased
because unboosted lobe emission dominates the whole spectrum at this
low frequency.

\subsection{The UV-bump}
SM89 fitted accretion disk spectra to a variety of quasars. They
collected IR to UV data from a sample of quasars -- most of them
belonging to the PG sample -- and used standard relativistic thin
accretion disk models including the effects of rotating black holes
for their fits of the `UV-bump'. They also assumed an underlying
powerlaw, which is no longer state of the art, but does not invalidate
neither their nor our conclusions although it might introduce a slight
and systematic shift in the luminosities.

{}From this detailed study we will mainly use one important number,
namely the {\it fitted} luminosity of the UV-bump. This number may
differ from the total integrated luminosity of the measured UV-bump, as
the fitted spectra do not always interpolate the observed data
perfectly well, indicating the presence of other components not
explained by and not necessarily linked to a standard accretion disk.
In column 6 (`UV') of Table 1 we listed this number from SM89 for
all objects, where we assumed an average inclination angle of
$40\degr$ for the disks.

Unfortunately, SM89 did not include all PG quasars in their
investigation, we therefore looked into the archived data of this
study and found a few more PG quasars with yet unpublished data. Using
exactly the same method as SM89, we were able to obtain some more disk
fits for these objects and included them in our list. The new sources
are indicated with an asterisk in Table 1: PG 0007+106, 0232$-$042,
0906+484, PG 1352+011, PG 1411+442, PG 1501+106, PG 1522+101, PG
1613+658, PG 1634+706, 2041$-$105, and PG 2302+029.  We will refer to
the whole sample of sources having UV-bump fits available (published
or unpublished) simply as the SM89 sample.

In the following we assume that the luminosities derived from the
UV-bump fits are indeed the `disk luminosity'. Because the PG sample
is so well studied we now have a large number of quasars having
disk fits and several other luminosity indicators available, such as
emission line luminosities and monochromatic continuum luminosities. As
it is generally accepted that these emission lines are produced
through ionization by the central source we can now compare UV-bump
fits and emission lines to get empirical conversion factors. For
sources where no fits are available we then can use monochromatic
luminosities and emission line luminosities to get estimates for the
disk luminosity from these numbers alone. For an individual source
there may be discrepancies between the 'real' value and the estimated
disk luminosity but in a statistical sense we will get reliable
numbers.

Of course this works only if the correlations are almost linear and
depend on one parameter only, i.e. the disk luminosity.  That this
is indeed the case is not obvious a priori. For example, an emission
line responds not only to an increase in the luminosity, but also to a
frequency shift of the exciting spectrum which in accretion theory
happens due to a shift in mass and accretion rate of the black hole.
Fortunately, the dependence of the shift in frequency ($T_{\rm eff}$)
is weak -- $T_{\rm eff}\propto \dot{M}_{disk}^{.25}/M_\bullet^{.5}$
while $L_{\rm disk}\propto\dot M_{\rm disk}$. If all quasars radiate
close to the Eddington luminosity, then we have $T_{\rm eff}\propto
M_\bullet^{-.25}$. In this sense the PG quasars are relatively uniform
-- they have comparable effective temperatures of several $10^4$
Kelvin -- and the expected relations should be fairly independent of
the black hole mass, but one has to remember that they will become
inadequate if $T_{\rm eff}$ changes appreciably. Some Seyfert galaxies
for example show a bump which peaks somewhere in the soft x-ray regime
and they should deviate systematically from such correlations. There
may also be plenty other (e.g. environmental) effects influencing the
emission line strength and one should be extremely carefull in
extrapolating the relations presented in the next section to other
luminosity regimes or even other quasar samples. The reason we can use
the simple conversion here is the uniformity of the PG quasars which
were all selected as objects with a blue excess -- but even there the
conversion from line luminosity to $L_{\rm disk}$ will be only correct
in a statistical sense and may be completely wrong for some individual
sources (see for example PG 1008+133 and PG 1206+459).

\subsection{Monochromatic continuum luminosities}
A very good indicator for the disk luminosity are the absolute blue
magnitude $M_{\rm b}$ -- which is not surprising as it was used to
select the quasars in this sample -- and the absolute magnitude
$M_{\rm v}$ of the continuum at $\lambda5500$ of the core spectrum in
the restframe of the quasar (BG92). We corrected the magnitudes given
in BG92 for the different value $q_0=0.5$ we use in our paper. The
flux $F_{\nu}$ at $\nu=5.45\cdot10^{14}$ Hz ($\lambda=5500$\AA) can be
converted to $M_{\rm v}$ by the formula
\begin{eqnarray}
M_{\rm v}&=&-2.5 \lg\left({ F_{\nu} \over 3.53\cdot 10^{-20} {\rm
erg/sec/Hz}}\right)\nonumber\\&&+5\lg\left({D_{l_\nu}\over {\rm pc}}\right)-5
\end{eqnarray}
and the monochromatic luminosity distance for our set of cosmological
parameters as a function of redshift is
\begin{equation}
D_{l_\nu}=1.2\cdot10^{10}\,{\rm pc}\;\left(\sqrt{1+z}-1\right).
\end{equation}

In Fig. 1a+b we plotted the relation between these colors and the disk
luminosity from SM89. In all cases the slope of the correlation
between color and luminosity found with linear regression is within
the errors 0.4 -- compatible with a direct proportionality between the
luminosity in the blue color and the total luminosity. We therefore
fixed the slope at that value and determined the intersection to a
high precision.
\begin{eqnarray}\label{l2mb}
\lg L_{\rm disk}\left(M_{\rm b}\right)&=&(-0.4 M_{\rm b}+35.90\pm0.04)\pm
0.29\\\label{l2mv}
\lg L_{\rm disk}\left(M_{\rm v}\right)&=&(-0.4 M_{\rm v}+35.70 \pm0.04)\pm
0.24.
\end{eqnarray}
The first error gives the uncertainty in the determination of the
offset and the second error the standard deviation of the sample.  As
one can see in Fig. 1 the correlation indeed is very tight and at
least for our sample we can expect that the disk luminosities inferred
from $M_{\rm b}$ with help of Eq.(\ref{l2mb}) will be close to the
fitted one. Two sources (PG 1008+133 \& PG 1206+459) showed extremely
high deviations (more than a factor 10 above the median deviation) in
the $L_{\rm disk}-M_{\rm b}$ correlation and we excluded them from
this fit. Their extreme deviation seems to indicate an intrinsic and
qualitative difference from the other sources and is probably not of
statistical origin. Further investigation of these two sources is
clearly indicated.

\subsection{Emission lines}
Another often used indicator for the disk luminosity is the luminosity
of the emission lines seen in the optical spectrum. BG92 published a
list of spectral data of all low redshift PG quasars and once again we
use the SM89 quasars as calibrators for the disk luminosity.  In Fig.
1c-e we show the relations between the luminosity in the emission
lines (He {\sc II}, {H$\beta$} and O {\sc III}) and the luminosity of
the disk. We make use of the definition in BG92 where for an emission
line EL with equivalence width EW the absolute magnitude of this line
is given by $M_{\rm [EL]}=M_{\rm v}-2.5\lg({\rm EW[EL]})$.

\begin{eqnarray}
\label{l2HeII} \lg L_{\rm disk}\left(M_{\rm He { II}}\right)&=&
\left(-0.4 M_{\rm He { II}}+{34.93\pm0.08}\right)\pm0.41\\
\label{l2Hbeta} \lg L_{\rm disk}\left(M_{{\rm H} {\beta}}\right)&=&\left(
-0.4M_{{\rm H} {\beta}}+{33.70\pm0.04}\right)\pm0.22\\
\label{l2OIII} \lg L_{\rm disk}\left(M_{\rm O { III}}\right)&=&
\left(-0.4 M_{\rm O { III}}+34.45\pm0.08\right)\pm 0.44
\end{eqnarray}

For our use it may be simpler but equally effective to use a total
luminosity of all lines and calculate a total emission line
luminosity. By adding all lines together
\begin{eqnarray}
M_{\rm lines}=M_{\rm v}-2.5\lg\left(EW[\mbox{He{\sc
II}}]\right.&+&EW[{\rm H} \beta]\nonumber\\&+&\left.EW[{\mbox{O {\sc
III}}}]\right)
\end{eqnarray}
we get (Fig. 1f)
\begin{equation}
\label{LEL} \lg L_{\rm disk}\left(M_{\rm lines}\right)=
\left(-0.4 M_{\rm lines}+33.60\pm0.04\right)\pm0.24.
\end{equation}

We found that the exact combination of the different line luminosities
(i.e. by using certain weights) did not really influence the result.
The good correlations we find show that we are not subject to severe
orientation dependent effects as far as the determination of the disk
luminosity is concerned although some of the line luminosities usually
are considered isotropic and others not. Nevertheless, H$_\beta$ shows
the tightest correlation and due to its large luminosity it dominates
$L_{\rm disk}\left(M_{\rm lines}\right)$. Comparing the correlation of the
emission lines with the $M_{\rm v}$ correlation one can see that there
is neither a tremendous gain nor a big loss in accuracy by using
emission line fluxes instead of continuum fluxes.  Both methods
therefore seem to be equally suitable to estimate the disk luminosity.

\begin{figure*}
\centerline{\bildh{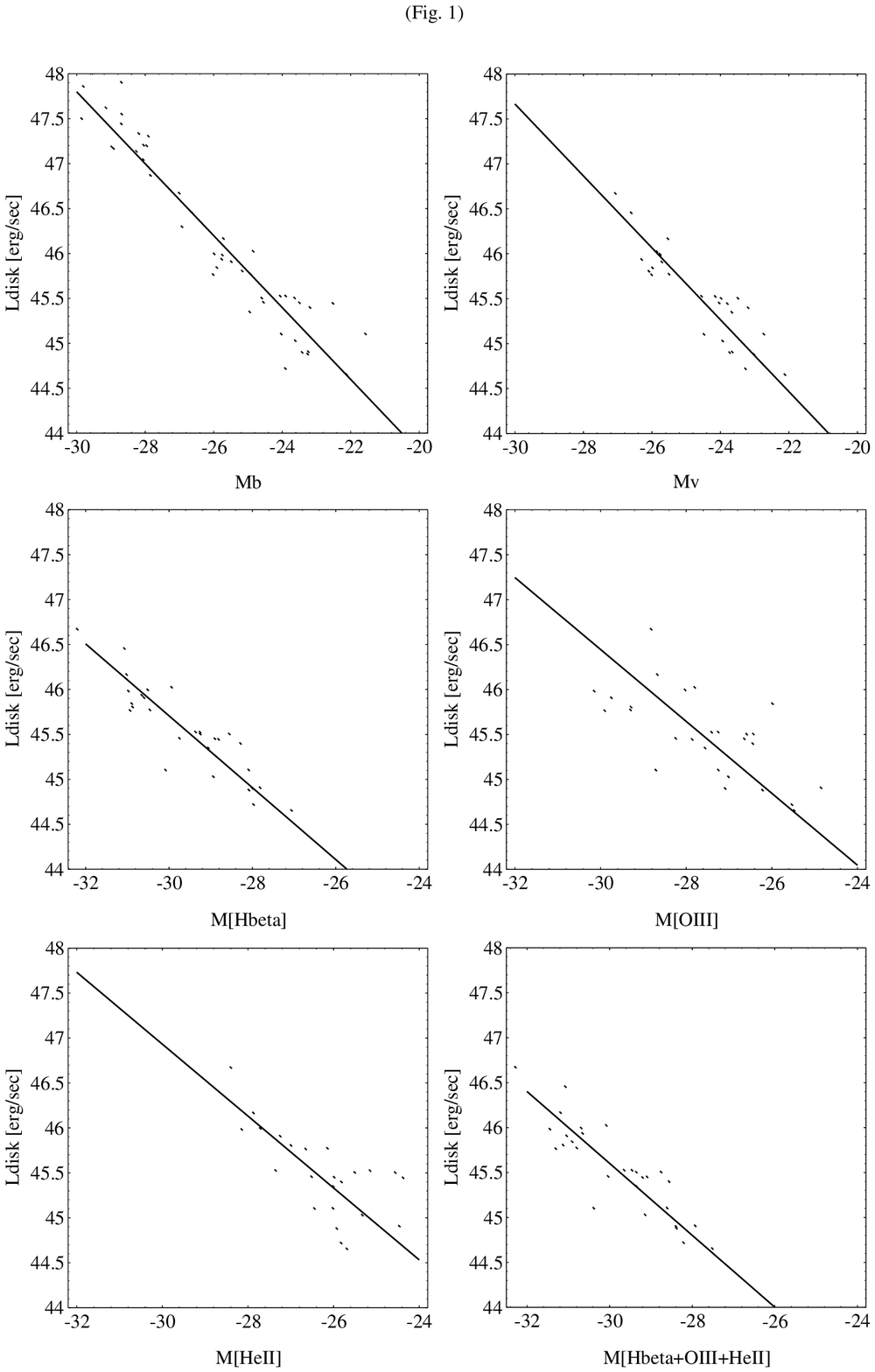}{0.9\textheight}{bbllx=4.3cm,bblly=2.4cm,bburx=19.3cm,bbury=23.9cm,clip=}}
\caption[]{Relation between the disk luminosity obtained from
accretion disk fits to the UV-bump by SM89 and (a) the
blue magnitude, (b) the visual magnitude and (c-e) emission line
luminosities ({He {\sc II}}, {H {\sc $\beta$}}, {O {\sc III}} and (f)
sum of all lines $M_{\rm lines}$) measured by Boroson \& Green (1992)
for PG quasars (from top left to bottom right).}
\end{figure*}

\subsection{Average disk luminosity}\label{averagedisk}
For the comparison between disk luminosity and radio emission, we make
use of an average disk luminosity obtained by taking the geometrical
mean from all available luminosity indicators (UV-bump, $M_{\rm v}$,
$M_{\rm b}$, $M_{\rm lines}$) of each source where the disk fit was
weighted by a factor three. The factor three was used to assure that
the fitted $L_{\rm disk}({\rm UV-bump})$ is not dominated by the
indirectly inferred numbers $L_{\rm disk}(M_{\rm b}),\,L_{\rm
disk}(M_{\rm v})$ and $L_{\rm disk}(M_{\rm lines})$ which in our
sample come in triplicate.  The geometric mean was choosen because the
derivation of the conversion factors and the subsequent discussion
takes place in the log/log plane -- the geometric mean is just the
arithmetic mean of the log.

As seen in Table 1 in most quasars all methods to derive the disk
luminosity yield similar results with a scattering smaller than a few
tenth in the log.  This confirms our optimism in anticipating that the
luminosity indicators are fairly reliable in a statistical sense.
Only a few exceptional cases show deviations larger than a factor
three.  These quasars, indicated with daggers in Table 1, have shown
large continuum variations between their photographic measurement in
the PG survey and the later spectrophotometry used for disk model
fitting.

\section{The radio data}
{}From observations we know that the radio spectrum of the compact cores
in radio loud quasars is flat and fairly independent of the
frequency. This is readily explained as the overlap of several
synchrotron spectra in a gradually declining magnetic field along the
jet cone (Blandford \& K\"onigl 1979).  Generally the observation of a
core at one frequency is equivalent with probing the jet at a certain
distance from the origin but as long as the spectrum is basically
flat, one frequency tells as much as any other.  In contrast to a
black body spectrum the shift of the flat non-thermal spectrum in
frequency would hardly make any difference in the observed flux.
Hence we do not need to integrate the whole spectrum to get a radio
luminosity in analogy to the UV luminosity.  For our comparison the
monochromatic luminosity at one radio frequency alone is adequate. In
{\em Paper I} we speculated that the compact radio emission in radio
weak quasars might be due to jets as well following the same basic
principles as their radio loud counterparts.  Consequently this and
the following considerations may apply to radio weak quasars as well.
Unfortunately only a few radio weak cores have been measured reliably
at high radio frequencies.  Antonucci \& Barvainis (1988) for example
find a high frequency excess in 9 out of 17 objects at $\lambda$2 cm
which could be attributed to a flat spectrum core.

The radio data we took are primarily from Miller et al. (1993) for the
low redshift PG sample and from Kellermann et al. (1989) for the rest
of the sample; they give a list of VLA fluxes at 5 GHz. 79 out of 89
quasars were detected and in 50 cases a central compact core
was also clearly seen. We added, if available, photometric radio data
from the NASA Extragalactic Database (NED), from a catalog of
Markarian galxies available at the MPIfR (Sherwood, W.  p.c.) and
other available literature (see Notes). We then transformed the
spectra into the rest frame of the quasars and fitted powerlaws or 3rd
order polynomials for each source. In cases where multiple photometric
datapoints were available at one frequency and the source was
variable, we took the geometric mean of the data.  For sources without
spectral informations we assumed an spectral index of $\alpha=-0.5$.
Thus we obtained fairly accurate total fluxes at 5 GHz in the rest
frame of the quasar. For the core fluxes we did the same, however, as
we have data only at one frequency we simply assumed a flat spectrum
($\alpha=0$) to get the 5 GHz rest frame flux. In the cases where the
extended emission swamps the core or the quasars were not detected at
all, we took the limits for the core fluxes or the limits of the total
fluxes respectively.  For a few sources among those not belonging to
the PG sample where no core fluxes were available, we either estimated
an upper limit for a flat spectrum core flux from the steep spectrum
at high frequencies or assumed a median value of the core/lobe ratio
of 0.72 found from our (radio weak) sample for Mkn 132, Mkn 679 and
1821+643 and of 0.15 (from double lobed sources) for 3C 110 and 3C
232.  We classified the radio spectra, which are shown in Falcke
(1994a), according to their slope at 5 GHz in the rest frame as flat
(F) or steep (S) and also marked obviously variable (V) spectra.
These classifications and the monochromatic core luminosities are
presented in Table 1 as well.
{\footnotesize
\begin{table*}
\begin{center}
\begin{tabular}{|llllllllllll|}
\hline
&&&&&\multicolumn{4}{c}{\hrulefill$L_{\rm
disk}$\hrulefill}&\multicolumn{3}{l|}{\hrulefill Radio at 5 GHz \hrulefill}\\
\#&name&$z$&class&$L_{\rm disk}$&UV&Mb&Mv&lines&$\nu L_\nu$&$f_{\rm
c}(\%)$&$\alpha$\\
\hline\hline
1&0003+158,PG&0.45&L,S,D&46.3&-&46.3&46.4&46.4&42.5&30.&-0.8\\
2&0003+199,PG,Mkn335&0.025&Q,P&44.6&44.6&44.7&44.5&44.6&38.6&80.&-0.29\\
3&0007+106,PG,IIIZw2$^*$&0.089&L,F,V&45.3&45.4&44.8&45.2&45.3&41.6&50.&0.62\\
4&0026+129,PG&0.142&Q,E&45.6&-&45.7&45.5&45.5&$<$39.2&$<$7.&-0.5\\
5&0043+039,PG&0.384&Q&46.&45.8&46.2&46.1&46.&39.7&90.&-0.5\\
6&0049+171,PG&0.064&Q,P&44.6&-&44.6&44.4&44.7&38.7&100.&-0.5\\
7&0050+124,PG,IZW1&0.06&Q,P&45.&44.9&45.2&45.2&45.&39.1&70.&-0.77\\
8&0052+251,PG&0.155&Q,E&45.5&-&45.6&45.5&45.5&39.3&60.&-0.5\\
9&0157+001,PG,MK1014&0.164&Q,P&45.6&-&45.7&45.5&45.6&40.4&60.&-1.\\
10&0804+761,PG&0.1&Q,E&45.4&-&45.3&45.4&45.5&39.3&40.&-0.5\\
11&0838+770,PG&0.131&Q&45.&-&45.1&45.&45.&$<$38.7&$<$90.&-0.5\\
12&0844+349,PG,TON 951&0.064&Q&44.9&44.7&45.4&45.&44.9&$<$38.4&$<$100.&-0.5\\
13&0921+525,PG&0.035&Q,E&44.3&-&44.2&44.2&44.5&38.7&50.&-0.87\\
14&0923+129,PG,Mrk705&0.029&Q,P&44.3&-&44.3&44.3&44.3&38.7&30.&-0.5\\
15&0923+201,PG&0.19&Q&45.5&-&45.5&45.5&45.6&$<$39.2&$<$90.&-0.5\\
16&0934+013,PG&0.05&Q,E&44.3&-&44.2&44.2&44.4&$<$38.2&$<$60.&-0.5\\
17&0947+396,PG&0.206&Q&45.4&-&45.4&45.3&45.4&$<$39.3&$<$90.&-0.5\\
18&0953+414,PG&0.239&Q,E&46.1&46.2&46.1&45.9&46.1&$<$39.2&$<$9.&-0.5\\
19&1001+054,PG&0.161&Q,E&45.3&-&45.3&45.3&45.2&$<$39.4&$<$60.&-0.5\\
20&1004+130,PG,4C13.41&0.24&L,S,D&45.9&46.&45.7&46.&45.6&41.4&5.&-1.\\
21&1011$-$040,PG&0.058&Q,P&44.6&-&44.7&44.7&44.5&38.3&100.&-0.5\\
22&1012+008,PG&0.185&Q,P&45.6&-&45.5&45.5&45.6&39.7&70.&-0.5\\
23&1022+519,PG&0.045&Q,E&44.2&-&44.2&44.2&44.1&38.&60.&-0.5\\
24&1048$-$090,PG&0.344&L,S,D&46.&-&46.&45.9&45.9&42.&7.&-0.88\\
25&1048+342,PG&0.167&Q&45.5&-&45.5&45.2&45.6&$<$39.&$<$90.&-0.5\\
26&1049$-$005,PG&0.357&Q,E&46.1&-&46.1&46.&46.1&$<$39.7&$<$60.&-0.5\\
27&1100+772,PG,3C249.1&0.313&L,S,D&45.9&45.9&46.&46.&46.&42.&8.&-0.99\\
28&1103$-$006,PG&0.425&L,S,D&46.&46.&46.2&46.&45.9&42.4&20.&-0.55\\
29&1114+445,PG&0.144&Q&45.1&45.&45.3&45.2&45.3&$<$38.9&$<$90.&-0.5\\
30&1115+407,PG&0.154&Q&45.1&-&45.3&45.1&44.9&$<$39.&$<$70.&-0.5\\
31&1116+215,PG&0.177&Q,P&45.9&-&45.8&45.9&46.1&40.&70.&-0.5\\
32&1119+120,PG&0.049&Q,E&44.7&-&44.9&44.7&44.5&$<$38.2&$<$30.&-0.5\\
33&1121+422,PG&0.234&Q&45.5&-&45.7&45.4&45.4&$<$39.2&$<$90.&-0.5\\
34&1126$-$041,PG&0.06&Q&44.8&-&44.7&44.9&44.9&$<$38.4&$<$70.&-0.5\\
35&1149$-$110,PG&0.049&Q,E&44.5&-&44.6&44.4&44.6&38.8&50.&-0.5\\
36&1151+117,PG&0.176&Q&45.5&45.5&45.6&45.3&45.3&$<$39.&$<$90.&-0.5\\
37&1202+281,PG,GQ COM&0.165&Q,P&45.4&45.3&45.8&45.1&45.3&39.5&80.&-0.5\\
38&1211+143,PG&0.085&Q&45.5&45.5&45.4&45.5&45.5&$<$39.&$<$60.&-0.5\\
39&1216+069,PG&0.334&Q,P&46.1&-&46.1&46.1&46.1&40.8&90.&-0.5\\
40&1226+023,PG,3C273&0.158&L,F,V&46.6&46.7&46.6&46.5&46.5&44.1&70.&-0.085\\
41&1229+204,PG,TON 1542&0.064&Q&44.9&44.9&45.1&44.9&45.&$<$38.7&$<$90.&-0.5\\
42&1244+026,PG&0.048&Q,E&44.1&44.&44.3&44.4&44.2&38.4&60.&-0.5\\
43&1259+593,PG&0.472&Q&46.4&46.5&46.5&46.3&46.&$<$39.8&$<$80.&-0.5\\
\hline\end{tabular}
\end{center}
 \caption[]{The quasar sample used in this paper. (\#1-\#87) is the PG
sample with $z<0.5$, (\#88-\#113) is the rest of the PG sample, and
(\#114-\#131) are quasars from the sample of SM89. Col. (1): ID
number, Col (2): IAU name; Col. (3): redshift, Col. (4): radio
classification Q -- radio weak, L -- radio loud, S -- steep spectrum
($\alpha\le-0.5$), F -- flat spectrum ($\alpha>-0.5$), V -- variable
spectrum, P -- point source, E - extended emission, D -- (double) lobe
structure; Col. (5): average disk luminosity; Col. (6-9): disk
luminosities derived from (6) UV-bump fits, (7) $M_{\rm b}$,
(8)continuum at $\lambda5500$ rest frame, (9) emission lines; Col.
(10): radio luminosity of the core at 5 GHz rest frame, Col. (11):
ratio of core flux to total flux at 5 GHz rest frame in per cent; Col.
(12): differential spectral index of fitted spectrum at 5 GHz rest
frame. An asterisk ($^*$) marks quasars with new accretion disk fits
and a dagger ($^\dagger$) marks the 2 quasars with strong variations
between $M_{\rm b}$ and the UV.}
\end{table*}}

{\footnotesize
\begin{table*}
\begin{center}
\begin{tabular}{|llllllllllll|}
\hline
&&&&&\multicolumn{4}{c}{\hrulefill$L_{\rm
disk}$\hrulefill}&\multicolumn{3}{l|}{\hrulefill Radio at 5 GHz \hrulefill}\\
\#&name&$z$&class&$L_{\rm disk}$&UV&Mb&Mv&lines&$\nu L_\nu$&$f_{\rm
c}(\%)$&$\alpha$\\
\hline\hline
44&1302$-$102,PG&0.286&L,F,V&46.1&-&46.2&46.3&45.8&43.&100.&0.035\\
45&1307+085,PG&0.155&Q&45.5&45.5&45.6&45.5&45.6&39.2&90.&-0.5\\
46&1309+355,PG&0.184&L,F,P&45.5&-&45.7&45.5&45.3&41.4&90.&-0.5\\
47&1310$-$108,PG&0.035&Q&44.3&-&44.2&44.2&44.5&$<$37.8&$<$100.&-0.5\\
48&1322+659,PG&0.168&Q&45.4&-&45.5&45.3&45.3&39.&90.&-0.5\\
49&1341+258,PG&0.087&Q&44.8&-&44.8&44.7&44.8&$<$38.5&$<$100.&-0.5\\
50&1351+236,PG&0.055&Q,E&44.4&-&44.5&44.6&44.1&$<$38.2&$<$40.&-0.5\\
51&1351+640,PG&0.087&L,F,P&45.2&-&45.1&45.3&45.2&40.7&100.&-0.14\\
52&1352+183,PG&0.158&Q&45.4&-&45.5&45.3&45.4&39.&90.&-0.5\\
53&1354+213,PG&0.3&Q&45.5&-&46.&45.3&45.2&$<$39.3&$<$90.&-0.5\\
54&1402+261,PG&0.164&Q,P&45.4&-&45.6&45.4&45.3&39.3&70.&-0.5\\
55&1404+226,PG&0.098&Q,P&44.8&-&45.&44.8&44.5&39.1&70.&-0.5\\
56&1411+442,PG$^*$&0.089&Q,P&45.3&45.5&45.3&45.1&45.1&38.9&90.&-0.5\\
57&1415+451,PG&0.114&Q&45.&-&45.2&45.&44.7&$<$38.8&$<$70.&-0.5\\
58&1416$-$129,PG&0.129&Q,E&45.3&45.1&45.4&45.5&45.8&39.4&20.&-0.5\\
59&1425+267,PG,TON 202&0.366&L,S,D&45.9&45.8&46.2&46.1&46.1&41.9&20.&-0.75\\
60&1426+015,PG,Mkn1383&0.086&Q,P&45.3&45.4&45.2&45.3&45.2&39.2&90.&-0.5\\
61&1427+480,PG&0.221&Q&45.4&-&45.5&45.2&45.5&$<$39.2&$<$90.&-0.5\\
62&1435$-$067,PG&0.129&Q&45.4&-&45.4&45.3&45.4&$<$38.8&$<$90.&-0.5\\
63&1440+356,PG&0.077&Q,E&45.&-&45.1&45.&44.9&38.8&40.&-1.1\\
64&1444+406,PG&0.267&Q&45.7&-&45.8&45.7&45.5&$<$39.2&$<$90.&-0.5\\
65&1448+273,PG&0.065&Q,P&44.9&-&45.&45.&44.7&38.9&100.&-0.5\\
66&1501+106,PG,Mkn841$^*$&0.036&Q,E&44.9&45.1&44.4&44.8&45.&38.3&50.&-0.81\\
67&1512+370,PG,4C37.43&0.371&L,S,D&46.&46.&46.1&46.&46.2&42.1&20.&-0.71\\
68&1519+226,PG&0.137&Q,E&45.2&-&45.2&45.1&45.1&$<$39.1&$<$30.&-0.5\\
69&1534+580,PG&0.03&Q,P&44.3&-&44.1&44.2&44.5&38.5&90.&-0.81\\
70&1535+547,PG&0.038&Q&44.5&-&44.4&44.5&44.6&38.1&100.&-0.5\\
71&1543+489,PG&0.4&Q,P&45.9&-&46.1&45.8&45.7&40.2&60.&-0.5\\
72&1545+210,PG,3C323.2&0.266&L,S,D&45.8&45.8&45.8&45.9&45.9&41.7&4.&-0.98\\
73&1552+085,PG&0.119&Q,E&45.&44.9&45.1&45.1&44.8&$<$38.9&$<$40.&-0.5\\
74&1612+261,PG&0.131&Q,E&45.3&-&45.2&45.2&45.6&39.7&30.&-0.5\\
75&1613+658,PG$^*$&0.129&Q,E&45.5&45.5&45.4&45.3&45.4&39.3&30.&-0.45\\
76&1617+175,PG&0.114&Q,E&45.2&-&45.3&45.2&45.3&$<$39.4&$<$50.&-0.5\\
77&1626+554,PG&0.133&Q,P&45.1&-&45.1&45.1&45.2&38.7&90.&-0.5\\
78&1700+518,PG&0.292&Q,E&46.&45.9&46.1&46.2&45.9&40.4&20.&-0.86\\
79&1704+608,PG&0.371&L,S,D&46.&-&46.1&46.2&45.8&41.1&0.5&-0.84\\
80&2112+059,PG&0.466&Q,P&46.5&-&46.5&46.5&46.5&40.3&70.&-0.5\\
81&2130+099,PG,IIZw 136&0.062&Q,E&45.2&45.4&45.1&44.9&45.&38.9&40.&-0.51\\
82&2209+184,PG&0.07&L,F,V&44.9&-&44.7&44.9&45.&41.2&100.&0.15\\
83&2214+139,PG&0.067&Q&45.&-&45.&45.&45.&38.3&100.&-0.5\\
84&2233+134,PG&0.325&Q,E&45.7&-&46.&45.7&45.5&39.7&60.&-0.5\\
85&2251+113,PG&0.323&L,S,D&45.9&45.8&45.9&46.1&46.1&41.5&3.&-0.72\\
86&2304+042,PG&0.042&Q,P&44.4&-&44.4&44.3&44.4&38.4&100.&-0.5\\
87&2308+098,PG&0.432&L,S,D&46.1&-&46.2&46.1&46.1&42.2&20.&-0.89\\
\hline\end{tabular}
\end{center}
(Table \thetable{} continued.)
\end{table*}
}

{\footnotesize
\begin{table*}
\begin{center}
\begin{tabular}{|llllllllllll|}
\hline
&&&&&\multicolumn{4}{c}{\hrulefill$L_{\rm
disk}$\hrulefill}&\multicolumn{3}{l|}{\hrulefill Radio at 5 GHz \hrulefill}\\
\#&name&$z$&class&$L_{\rm disk}$&UV&Mb&Mv&lines&$\nu L_\nu$&$f_{\rm
c}(\%)$&$\alpha$\\
\hline\hline
88&0044+030,PG&0.624&L,S&46.6&-&46.6&-&-&42.1&40.&-0.82\\
89&0117+213,PG&1.49&Q&47.3&-&47.3&-&-&$<$40.5&$<$60.&-0.5\\
90&0946+301,PG&1.22&Q&47.1&47.1&47.1&-&-&$<$40.1&$<$70.&-0.5\\
91&1008+133,PG$^\dagger$&1.29&Q&47.3&47.7&46.3&-&-&$<$40.2&$<$70.&-0.5\\
92&1115+080,PG&1.72&Q&47.6&47.6&47.5&-&-&$<$40.3&$<$60.&-0.5\\
93&1138+040,PG&1.88&Q&47.4&47.4&-&-&-&$<$40.3&$<$60.&-0.5\\
94&1148+549,PG&0.969&Q&47.1&47.2&47.&-&-&40.7&40.&-0.5\\
95&1206+459,PG$^\dagger$&1.16&Q&45.6&45.1&47.2&-&-&$<$40.&$<$70.&-0.5\\
96&1222+228,PG,TON 1530&2.05&Q&47.6&47.5&47.7&-&-&42.2&50.&-0.5\\
97&1241+176,PG&1.27&L,S/F&47.2&47.2&47.4&-&-&43.2&50.&-0.50\\
98&1247+267,PG&2.04&Q&47.8&47.9&47.7&-&-&41.1&40.&-0.5\\
99&1248+401,PG&1.03&Q&46.9&46.9&46.9&-&-&$<$40.1&$<$70.&-0.5\\
100&1254+047,PG&1.02&Q&47.&47.&47.&-&-&40.3&70.&-0.5\\
101&1329+412,PG&1.93&Q&47.2&47.2&47.4&-&-&40.5&50.&-0.5\\
102&1333+176,PG&0.554&L,?&46.6&-&46.6&-&-&41.9&80.&-0.5\\
103&1338+416,PG&1.22&Q&47.3&47.3&47.1&-&-&$<$40.1&$<$70.&-0.5\\
104&1352+011,PG$^*$&1.12&Q&47.2&47.2&47.&-&-&$<$40.1&$<$70.&-0.5\\
105&1407+265,PG&0.944&Q&47.&47.&-&-&-&41.3&30.&-0.5\\
106&1522+101,PG$^*$&1.32&Q&47.7&47.9&47.3&-&-&40.4&60.&-0.5\\
107&1538+477,PG&0.77&L,S&46.7&-&46.7&-&-&42.1&40.&-1.\\
108&1630+377,PG&1.47&Q&47.4&47.4&47.3&-&-&$<$40.2&$<$60.&-0.5\\
109&1634+706,PG$^*$&1.33&Q&47.9&48.&47.6&-&-&41.1&40.&-0.5\\
110&1715+53,PG&1.92&Q&47.4&-&47.4&-&-&40.9&40.&-0.5\\
111&1718+481,PG&1.08&L&47.5&47.5&47.3&-&-&42.9&50.&-0.5\\
112&2302+029,PG$^*$&1.04&Q&47.2&47.3&47.&-&-&40.2&40.&-0.5\\
113&2344+092,PG&0.677&L,F,V&46.4&46.3&46.6&-&-&43.8&90.&-0.18\\\hline\hline
114&0134+329,3C48&0.367&L,S&45.7&45.7&-&-&-&41.8&0.5&-0.94\\
115&0232$-$042,PHL1377$^*$&1.44&L,S&47.3&47.3&-&-&-&43.4&20.&-0.77\\
116&0237$-$233,PKS,MRC&2.22&L,S&47.5&47.5&-&-&-&$<$43.8&$<$5.&-0.15\\
117&0405$-$123,PKS&0.574&L,S&46.9&46.9&-&-&-&$<$43.2&$<$20.&-0.52\\
118&0414$-$060,3C110&0.781&L,S&46.6&46.6&-&-&-&42.4&10.&-0.75\\
119&0710+458,Mkn376&0.056&Q&44.4&44.4&-&-&-&39.&7.&-0.5\\
120&0742+318,4C31.30&0.462&L,F&46.3&46.3&-&-&-&43.3&70.&-0.23\\
121&0906+484$^*$&0.118&Q&45.6&45.6&-&-&-&$<$39.&$<$90.&-0.5\\
122&0955+326,3C232&0.533&L,S&46.2&46.2&-&-&-&42.6&10.&-0.53\\
123&0958+551,Mkn132&1.76&Q&47.7&47.7&-&-&-&$<$41.5&$<$40.&-0.5\\
124&1011+250,Ton490&1.63&L,F,V&46.9&46.9&-&-&-&43.9&100.&0.34\\
125&1317+277,Ton153&1.02&Q&47.&47.&-&-&-&$<$40.6&$<$70.&-0.5\\
126&1421+330,Mkn679&1.9&Q&47.4&47.4&-&-&-&41.7&40.&-0.5\\
127&1435+638&2.06&L,F,V&47.4&47.4&-&-&-&44.2&80.&-0.17\\
128&1821+643,E1821+643&0.297&Q&46.6&-&46.6&-&-&41.1&60.&-0.5\\
129&2041$-$109,Mkn509$^*$&0.036&Q&45.&45.&-&-&-&38.8&30.&-0.56\\
130&2145+067,4C06.69&0.99&L,F,V&46.8&46.8&-&-&-&44.4&100.&0.029\\
131&2201+315,4C31.63&0.297&L,F,V&46.1&46.1&-&-&-&43.4&80.&-0.025\\
\hline\end{tabular}
\end{center}
(Table \thetable{} continued. The extended sample.)
\end{table*}
}

\begin{table*}
\begin{center}
\begin{tabular}{|lllllll|}
\hline
&&&\multicolumn{1}{c}{\hrulefill$L_{\rm
disk}$\hrulefill}&\multicolumn{3}{l|}{\hrulefill Radio at 5 GHz \hrulefill}\\
\#&name&$z$&$f\cdot Q_{\rm jet}$&$\nu L_\nu({\rm core})$&$f_{\rm
c}(\%)$&$\alpha$\\
\hline\hline
132&0106+13,3C33&0.059&45.4&40.2&0.6&-0.77\\
133&0109+49,3C35.0&0.066&44.4&40.1&2.&-0.85\\
134&0152+28,3C42&0.395&45.9&40.8&0.2&-0.9\\
135&0132+37,3C46&0.437&46.&40.7&0.5&-0.87\\
136&0221+27,3C67&0.31&45.7&42.6&20.&-0.91\\
137&0307+16,3C79&0.255&46.2&41.&0.6&-1.1\\
138&0356+10,3C98&0.03&44.9&39.2&0.2&-0.65\\
139&0410+11,3C109&0.307&46.2&42.6&20.&-0.7\\
140&0411+14,4C14.11&0.38&45.1&41.7&2.&-0.77\\
141&0433+29,3C123&0.218&46.1&42.&0.6&-0.89\\
142&0702+74,3C173.1&0.292&46.1&41.&0.7&-1.\\
143&0734+80,3C184.1&0.119&45.&40.2&0.5&-1.3\\
144&0745+56,4C+56.16&0.035&44.4&40.7&20.&-0.98\\
145&0802+24,3C192&0.06&45.1&39.6&0.3&-0.69\\
146&0917+45,3C219&0.174&45.4&41.4&2.&-1.\\
147&0936+36,3C223&0.137&45.7&40.5&0.7&-0.78\\
148&0944+73,4C73.08&0.058&45.2&40.1&3.&-0.93\\
149&0958+29,3C234&0.185&45.9&41.9&7.&-0.97\\
150&1003+35,3C236&0.099&45.4&41.9&30.&-0.65\\
151&1030+35,3C244.1&0.428&45.9&$<$40.2&$<$0.04&-1.\\
152&1308+27,3C284&0.239&45.4&40.5&0.7&-1.1\\
153&1319+42,3C285&0.08&44.4&39.9&0.8&-0.82\\
154&1409+52,3C295&0.461&46.2&41.4&0.1&-1.1\\
155&1420+19,3C300&0.27&45.4&40.9&0.5&-0.99\\
156&1441+52,3C303&0.141&45.1&41.7&10.&-0.61\\
157&1522+54,3C319&0.192&45.&$<$39.8&$<$0.1&-1.\\
158&1529+24,3C321&0.096&45.4&40.7&2.&-0.92\\
159&1550+20,3C326&0.098&45.4&40.4&3.&-2.1\\
160&1626+27,3C341&0.448&46.7&40.3&0.1&-1.\\
161&1658+47,3C349&0.205&45.7&41.2&2.&-0.81\\
162&1832+47,3C381&0.161&45.7&40.4&0.3&-0.8\\
163&1833+32,3C382&0.058&44.9&41.1&9.&-0.65\\
164&1842+45,3C388&0.092&44.9&41.&3.&-0.93\\
165&1845+79,3C390.3&0.057&45.1&41.3&8.&-0.87\\
166&1939+60,3C401&0.201&45.4&41.4&2.&-1.\\
167&2141+27,3C436&0.215&45.4&41.1&1.&-1.\\
168&2153+37,3C438&0.29&45.9&41.1&0.4&-1.2\\
169&2243+39,3C452&0.082&45.2&41.2&4.&-0.87\\
170&2309+18,3C457&0.31&46.4&40.6&0.5&-0.91\\
\hline\end{tabular}
\end{center}
 \caption[]{The FR\,II sample used in this paper. Col. (1): ID
number, Col (2): IAU name; Col. (3): redshift, Col. (4): disk luminosity,
derived by multiplying the lobe power calculated by RS91 with a factor 10; Col.
(5): radio luminosity of the core at 5 GHz rest frame, Col. (6):
ratio of core flux to total flux at 5 GHz rest frame in per cent; Col.
(7): differential spectral index of fitted spectrum at 5 GHz rest
frame.}
\end{table*}

\section{The UV-radio correlation}
In Fig. 2 we plotted the 5 GHz radio core flux versus the average
disk luminosity determined from all available luminosity indicators
for each quasar (Sec. \ref{averagedisk}).
In Fig. 2 radio cores of radio weak quasars are represented by black
dots and an arrow indicates upper limits for the detection of
a compact core. In the few cases where the quasar was not detected at
all, we used only the arrow. Radio weak quasars showing extended
emission are marked with a grey shade around the black dot.

Cores of radio loud quasars are given by open circles and boxes. Point
sources are marked with a `P', double lobed sources with a `D' and
flat spectrum sources by a grey filling.

To complete the picture, we also added the sources belonging to the
extended sample in grey. Those sources belonging to the high redshift
PG quasar sample are marked with an additional cross.

\begin{figure*}
\centerline{\bildh{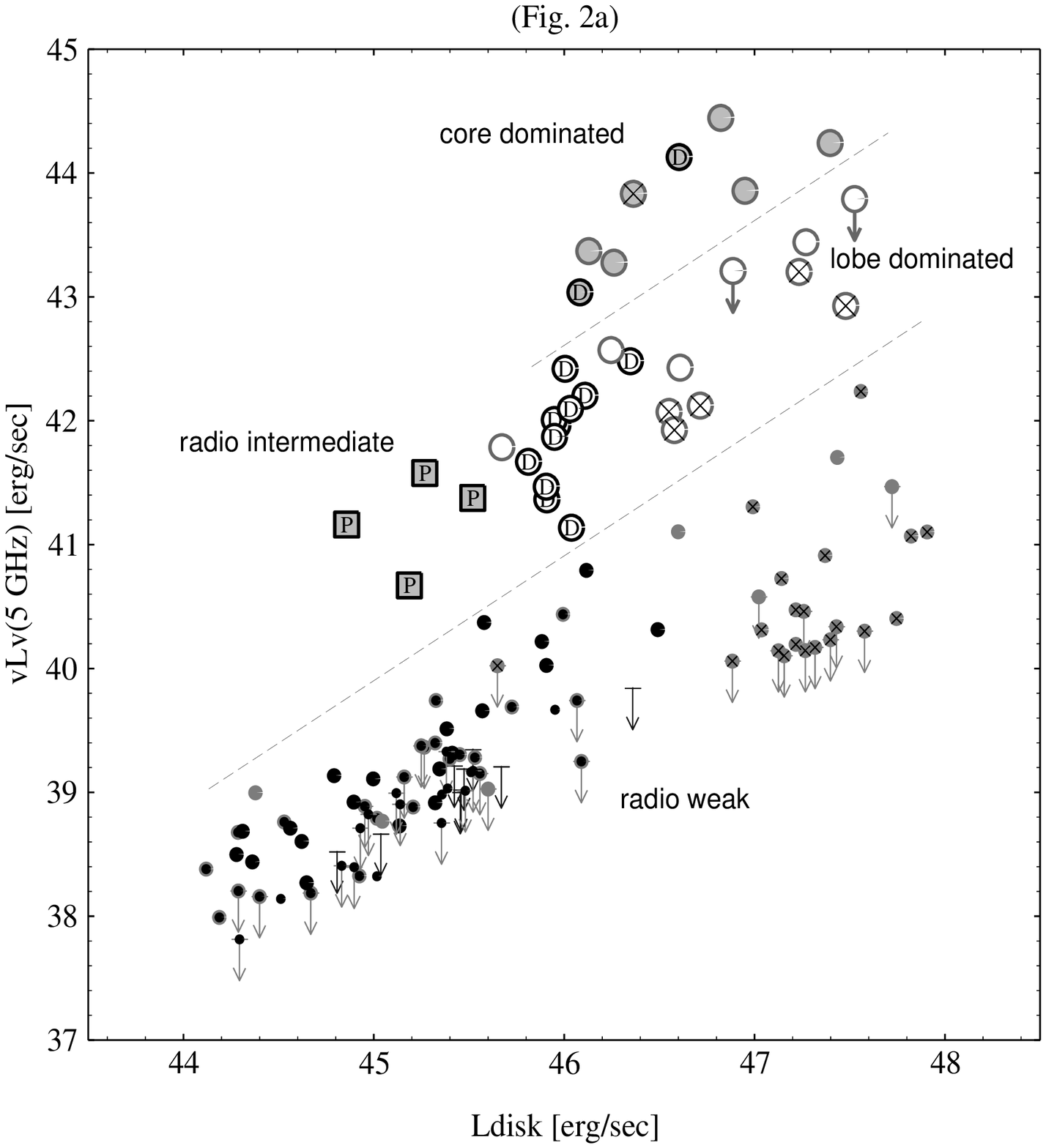}{17cm}{bbllx=1.5cm,bblly=6cm,bburx=20.1cm,bbury=25.2cm,clip=}}
\caption[]{Monochromatic luminosity of the compact radio core vs. the
disk luminosity (see text) for PG quasars with $z<0.5$. 'D': PG
quasars showing lobes, 'P': radio loud PG quasars appearing as point
sources, {\em boxes}: RIQ, {\em black dots}: radio weak point sources,
{\em dots with shades}: radio weak sources with extended emission,
{\em grey dots with cross}: radio weak PG quasars with $z>.5$, {\em
grey dots without cross}: radio weak quasars from SM89, {\em grey open
circles with cross:} radio loud PG quasars with $z>0.5$, {\em grey
open circles without cross:} radio loud quasars from SM89, {\em
circles and boxes with grey background:} flat spectrum sources.}
\end{figure*}

The quasars fall in at least three distinct categories: radio weak,
radio loud lobe dominated quasars (LDQ) and flat spectrum core
dominated quasars (CDQ).  The distinction between radio-loud and
radio-weak holds not only for the total flux, but also for the core
fluxes.  The separation needs to be undestood relative to the disk
luminosities.  There may be radio weak quasars with a high accretion
rate which have more luminous radio cores than some low accreting
radio loud objects.  Considering only the core fluxes, CDQ are the
brightest population. This diagram confirms the well accepted idea
that flat spectrum CDQ are the consequence of orientation effects in
conjunction with relativistic boosting, which leads to a strong
enhancement of the core flux relative to the un-boosted lobes and disk
luminosities.

A few sources do not fit into this general trend: the flat spectrum
variable point sources with no sign of large scale lobe structure and
a relatively low accretion rate with $L_{\rm disk}<10^{46}$ erg/sec
(boxes in Fig. 2).  They have overall properties of (boosted) CDQ but
are relatively less luminous and rather seem to fall on the
extrapolation of LDQ.  Following MRS, we will call these sources
radio-intermediate quasars (RIQ) and discuss them in more detail
later.

There is a clear correlation between the disk luminosity and the radio
core fluxes in each subgroup. Despite the many uncertainties involved,
this correlation is remarkably good for extragalactic astrophysical
data but not surprising; radio power an NLR luminosity are also well
correlated in these galaxies (see Wilson 1992).  It is noteworthy that
the upper limits also agree well with the general trend so that a
non-detection does not necessarily mean that there is no radio
emission, but that the sensitivity was insufficient to measure a radio
jet which belongs to a quasar with this redshift and accretion rate.
Our theoretical arguments imply that they should be detectable with
higher sensitivity (below the $100\mu$Jy level).  All correlations
remain when fluxes, rather than luminosities, are plotted.

We stress that the radio cores of radio loud quasars (i.e. LDQ and
CDQ) are brighter than the cores of radio weak quasars at the same
disk luminosity. As we know from numerous VLBI observations that the
bulk of the core emission in radio loud quasars comes from a very
compact region on the parsec scale it is evident that the difference
between radio loud and radio weak quasars is established already at
this small scale.

Finally we note that radio loud quasars with lobe emission
(morphologically similar to FR\,II radio galaxies) exist only at disk
luminosities $L_{\rm disk}\ga10^{46}$ erg/sec although radio weak
sources with 100 times lower accretion rate do exist in large numbers.
Unless there is a strange selection effect -- this is an optically
selected sample -- it means that disks must have luminosities larger
than a threshhold luminosity of $\sim10^{46}$ erg/sec to produce a
quasar with FR\,II type radio jet.

\section{Confronting theory with data}
\subsection{Scaling with disk luminosity and velocity}
What kind of correlation do we expect between radio core and UV
emission?  If we base our answer on the idea of a jet-disk symbiosis
than we find from {\em Paper I} that for a statistical sample the
radio flux scales with disk luminosity, bulk Lorentz factor
$\gamma_{\rm j}$, and Doppler factor ${\cal D}$ as
\begin{equation}
F({\rm radio\,core})\propto {\cal D}^{2.2}\gamma_{\rm j}^{-1.8}L_{\rm
disk}^{1.42}
\end{equation}
for radio loud and radio weak quasars.

There are a few more parameters involved like the relativistic
electron number density $x_{\rm e}$ in units of the total (proton)
particle density and the mimimum energy of the electrons
$\gamma_{\min,e}m_{\rm e}c^2$, the total jet power $q_{\rm j/l}$
normalized to the disk luminosity and the inclination angle $i$. For
homogenous classes of objects we assume that all these parameters will
scatter around a constant value. This will lead to a certain scatter
of the radio luminosity around an average value. To simplify the
analysis we will, however, assume that the dominant reason for
scattering will be the randomly oriented inclination angle with the
consequence of strong Doppler boosting. This can lead to a broadening
of the distribution by a factor 10-100 (see Fig.3). The assumption now
is that the scatter introduced by the other parameters is less than
this in the regime of relativistic flows. If this is too much a
simplification and the broadening of the distribution is caused by
intrinsic parameter variations than the average bulk Lorentz factors
we derive here have to be regarded as upper limits.  This, however,
seems unlikely as the Lorentz factors of 3-10 we obtain are quite
consistent with many other studies in the past (eg. Ghisellini et al.
1993).

In the earlier discussions on the nature of the quasar radio emission
it was suggested that relativistic beaming was responsible for the
scatter of all quasars (radio loud and radio weak) but as Strittmatter
et al. (1980) showed this is inconsistent with the flux density
distribution of optically selected quasar samples and there is a real
dichotomy in the radio flux. Hence, we suggest that the scatter within
each population is dominated by relativistic boosting and there is an
additional discrete process producing an offset between radio loud and
radio weak.

The scaling over several orders of magnitudes in the radio luminosity
thus will reflect mainly the change in the accretion rate which indeed
may vary over 4 orders of magnitudes in the quasars alone and even
more if one later includes Seyfert galaxies and stellar mass black
holes. The non-linearity between disk luminosity and radio flux is a
consequence of the non-linear response of the synchrotron emissivity
to a change in the magnetic energy density. Nevertheless, there still
remains one uncertainty in the discussion of the scaling, namely, we
have no idea whether the jet velocity also depends on the mass
accretion rate.  We only can argue that this dependence is not very
strong because the observed Lorentz factors in jets are usually
estimated to be in the range 3-30 (Ghisellini et al.  1993) with a
mean value of $\sim10$, despite a large range of mass accretion rates
of at least 2 orders of magnitudes for radio loud quasars.  One could
also argue that the escape velocity around black holes is a
substantial fraction of the speed of light, so that jets with speed
$\ll0.5 c$ from black holes are difficult to achieve unless there is a
perfect fine-tuning in the acceleration mechanism.  On the other hand
will the Compton drag of disk photons decelerate the jet if it starts
at $\gamma_{\rm j}\gg10$ (Melia \& K\"onigl 1989). In light of this
discussion we make the Ansatz that the bulk proper velocity of the jet
depends on the accretion rate and thus on the disk luminosity in the
quasar luminosity range\footnote{One should understand this scaling in
statistical terms. If the jet velocity depends on the ratio $L_{\rm
disk}/L_{\rm edd}$ and the average of this ratio has a slight
dependence on the total power (see SM89) this could produce such a
scaling indirectly.}  as
\begin{equation}\label{gammal46}
\gamma_{\rm j}\beta_{\rm j}=\gamma_{\rm j,0}\beta_{\rm
j,0}\left({L_{\rm disk}\over 10^{46}{\rm erg/sec}}\right)^\xi
\end{equation}
with the additional assumption that for quasars $\left|\xi\right|$ is
a small number ($<0.5$), $\beta_{\rm j,0}\simeq1$ and $\beta_{\rm
j}\simeq1$ (see above). Following our general philosophy that disks
and jets are basically the same everywhere we assume the same scaling
for radio weak quasars.

\subsection{Obscuring dust torus}
The PG quasars we use are selected by their blue color and not by
their radio flux. In our context this part of the spectrum is produced
by the accretion disk and should be relatively free of any substantial
selection effect like boosting. We only have to take into account
that an obscuring dust torus surrounding the nucleus could prevent us
from seeing edge on disks. This, on the other hand, is even
helpful because the disk luminosity is then even less dependent on
the inclination angle if we deal with maximally rotating black holes
as central objects (Cunningham 1975): light bending and a moderate
amount of boosting due to the high rotational velocity compensate the
effects of decreasing apparent surface with higher inclination angles.

It is not really known what the real opening angle of the viewing cone
permitted by the torus for each quasar population is and it might
iself depend on other parameters (i.e. jet power and $z$, see Kapahi
1990, Falcke et al. 1994). We pick for the assumed obscuring dust
torus of radio weak quasars deliberately a viewing cone with
semi-opening angle of $60\degr$ -- saying that we do not detect 50\%
of the quasars because of obscuration -- and obtain as an average
inclination angle for the jet axis with respect to the line of sight
in this sample $i\approx40\degr$. In the range $60\degr>i>18\degr$ we
should find $90\%$ of our sample. This value for the obscuring torus
is taken from McLeod \& Rieke (1994) who investigated the host
galaxies of a subsample of the PG quasars.

Studies of the ratio between FR\,II galaxies and radio loud quasars
suggest that this torus might there even cover a larger area than for
radio weak quasars and we take $45\degr$ as a default value (Barthel
1989). Which means, we miss more than 2/3 of all sources and the 90\%
dividing line is at 14\degr.  None of our conclusions depends
sensitively on these assumptions and we will later show that those
numbers are able to explain the UV-radio correlation in a consistent manner.

\subsection{Parameter space for radio loud jets}
We now discuss the range of parameters needed to explain the UV-radio
correlation as being due to a link between jet and disk. In {\em Paper
I} we presented the most simple minded model for the jet emission and
its parameters and noted that in order to explain the high radio flux
of radio loud cores we get close to the limits imposed by energy and
mass conservation in a jet-disk system.  We will fit the model step by
step to the data and discuss the range of possible parameters. Among
all possible models only the most efficient case, a `total
equipartition jet' --- implying rough equipartition between energy
stored in magnetic field, electrons and protons and bulk kinetic
motion --- is able to explain these radio fluxes. Let us now first fix
the plasma parameters of this type of jet such that they yield maximum
efficiency (see {\em Paper I}) which implies:
\begin{itemize}
\item[a)] a minimum Lorentz
factor of the relativistic electron distribution of $\gamma_{\rm
e}\approx100$ for a powerlaw distribution with $p=2$,
\item[b)] a relativistic electron fraction $x_{\rm e}\approx1$ which means
that in a proton/electron jet all electrons get accelerated or that
for every thermal electron one additional secondary particle (electron
or positron) is created and
\item[c)] a proton/electron ratio of $\mu_{\rm p/e}\approx2$ meaning equal
energy
in relativistic proton and electron distribution ($u_3\approx k_{\rm
e+p}\approx1$ and $\ln(\gamma_{\rm e,max}/\gamma_{\rm
e,min})\approx3.6$)
\end{itemize}
According to the discussion in {\em Paper I} one may vary some of
these `microscopic' parameters but there is no way to further increase
the emitted flux of the jet other than by changing the `macroscopic'
parameters which are
\begin{itemize}
\item[a)] the jet Lorentz factor $\gamma_{\rm j}$, assumed to be of the
order 10 ($\Rightarrow\beta_{\rm j}=1$),
\item[b)] the disk luminosity $L_{46}$ in units of $10^{46}$ erg/sec
which varies in our sample from $10^{-2}$ to $10^2$ and
\item[c)] the total jet power $q_{\rm j/l}$ in units of the disk
luminosity which we consider to be $<1$.
\end{itemize}
These parameters are fixed initially at these extreme values only to
formally describe the radiatively most efficient state of the jet but
they are not necessarily unphysical. For example if the electrons
responsible for the synchrotron emission are secondary pairs created
in hadronic cascades than a low energy cut-off at $\gamma_{\rm
e}\ga180$ is a natural consequence of the Pion decay and because one
is not limited by the number of thermal plasma electrons (as in a pure
shock acceleration scenario for the electrons) the ratio $x_{\rm e}$
of relativistic to thermal electrons may be very high.  The product
$x_{\rm e}\gamma_{\rm e,min}$ is then only limited by the total energy
constraints. When the parameters are fixed in this way and one
calculates the emitted flux for a supersonically expanding jet with
maximal internal energy contents one gets ({\em Paper I})
\begin{eqnarray}\label{lnujmax}
\nu L_\nu({\rm5 GHz})&=&6.7\cdot10^{42}\,{\rm erg/sec}\;
{\cal D}(L_{46},i_{\rm obs})^{2.17}\sin i_{\rm obs}^{0.17}\nonumber\\&&
\cdot\left({\gamma_{\rm e,min}x_{\rm e}\over 100}\right)^{0.83}
\left({6\over\gamma_{\rm j,0}}\right)^{1.8} q_{\rm j/l}^{1.42}
L_{46}^{1.42-\xi}.
\end{eqnarray}
where we have already made use of the possible velocity scaling (Eq.
\ref{gammal46}) and the parameter setting discussed above. The Doppler
factor ${\cal D}$ is defined as
\begin{equation}
{\cal D}=1/\gamma_{\rm j}(1-\beta_{\rm j}\cos i_{\rm obs}).
\end{equation}
For our comparison with the data we will assume that all jets are
two-sided.

In order to fix the parameters we first try to fit the cores of the
lobe dominated radio loud quasars at $L_{\rm disk}\simeq10^{46}$
erg/sec and find that already there the possible parameter range is very
narrow. The quasars can not be fitted with bulk Lorentz factors
$\gamma_{\rm j,0}\ga10$ and $q_{\rm j/l}<1$ as with increasing Lorentz
factor the radiative efficieny is decreased and the boosting effect
reduces the visible radio flux for the bulk of the population even
further. Together this makes a very strong effect which depends on the
Lorentz factor roughly as $\gamma_{\rm j,0}^{-4}$.

We also can get an approximate lower limit for the Lorentz factor from
our prerequisite that the width of the distribution is mainly given by
the scatter in inclination angles. A small Lorentz factor would yield
only a narrow band for the scattering width of the UV-radio
correlation and some sources would fall outside this band so we
conclude that $\gamma_{\rm j,0}\ga3$ ($q_{\rm j/l}\ga0.05$).  The
width (and therefore the scattering) of the radio flux distribution is
represented
in Fig. 3 by grey lanes. Quasars with larger inclination angle (below
the lane) are assumed to be missing in our sample due to an obscuring
dust torus and smaller inclination angles (above the lane) will lead
to a strong boosting effect hence $\sim 10\%$ of each sample should
lie close to or above this lane.

In a second step we include the remaining sources and first adopt the
velocity gradient by varying $\xi$ in Equation (\ref{gammal46}). As
noted before the gradient must not be very steep because of the fourth
power entering in Equation (\ref{lnujmax}); this is highlighted in
Fig. 3 where we show the expected UV-radio correlation for $\xi$ as
large as $0.5$. It is not possible to determine this evolution in more
detail from the few radio loud sources in our sample, especially as
the high power sources do not belong to the well defined subsample of
PG quasars with $z<0.5$. We only can hope that the distribution of the
high power sources is not completely misleading and estimate that
$0\la\xi\la0.25$.

Finally we include the CDQ in our analysis. A strict upper limit of
the radio flux as a function of disk luminosity is given by those
models with maximum boosting ($i=0$) and no source in our diagram
should be far above this line. This condition gives us additional
limits on the Lorentz factor and also $\xi$. Moreover should most of
the CDQ cluster around a line representing sources at the boosting
cone thus inclined to line of sight by $i\simeq1/\gamma_{\rm j}$. This
restricts the margins for the Lorentz factor even further. Figure 3
shows different parameter combinations and our best fit model with
estimated errors.
\begin{equation}
\gamma_{\rm j}=6^{+2}_{-2},\; q_{\rm j/l}=0.15^{+.2}_{-.1},\;
\xi=0.15^{+0.1}_{-0.15}
\end{equation}

The best fit parameter set implies that indeed the microphysical state
of the plasma must not be too far away from the radiatively most
efficient state, i.e. $x_{\rm e}\gamma_{\rm min,e}\approx100$,
otherwise we will have to increase $q_{\rm j/l}$ substantially.
Accordingly we have an indirect argument that the electron
distribution in radio loud quasars either must have a low energy
cut-off or consists of additionally created pairs. We suggested in
{\em Paper I} that both effects might naturally arise if the electron
population is dominated by pairs produced in the $\pi-$decay of
hadronic cascades initiated by $pp$ or $p\gamma$ collisions.

\begin{figure*}
\centerline{
\bildh{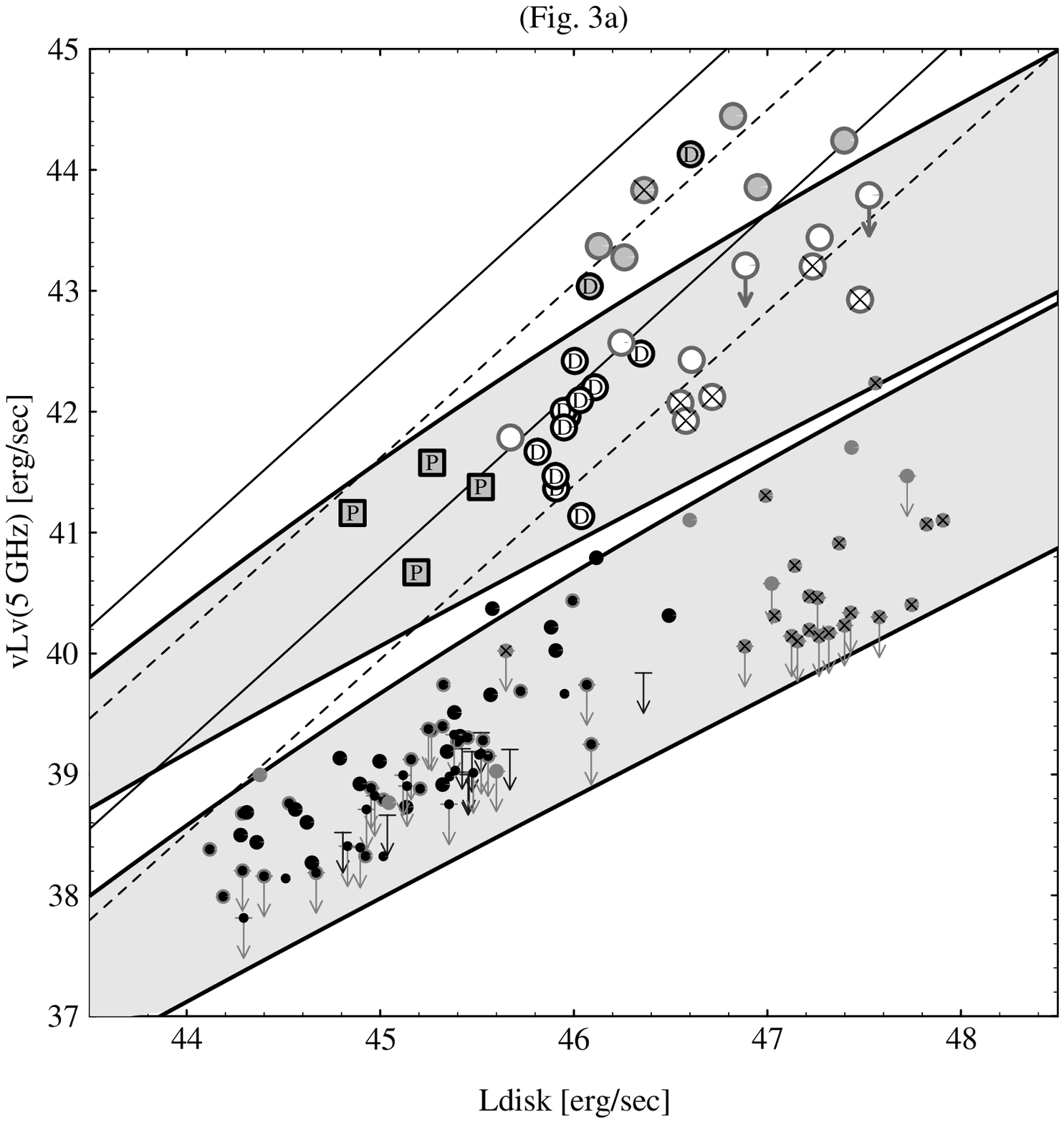}{8.5cm}{bbllx=2.7cm,bblly=5.3cm,bburx=20.0cm,bbury=22.2cm,clip=}
\bildh{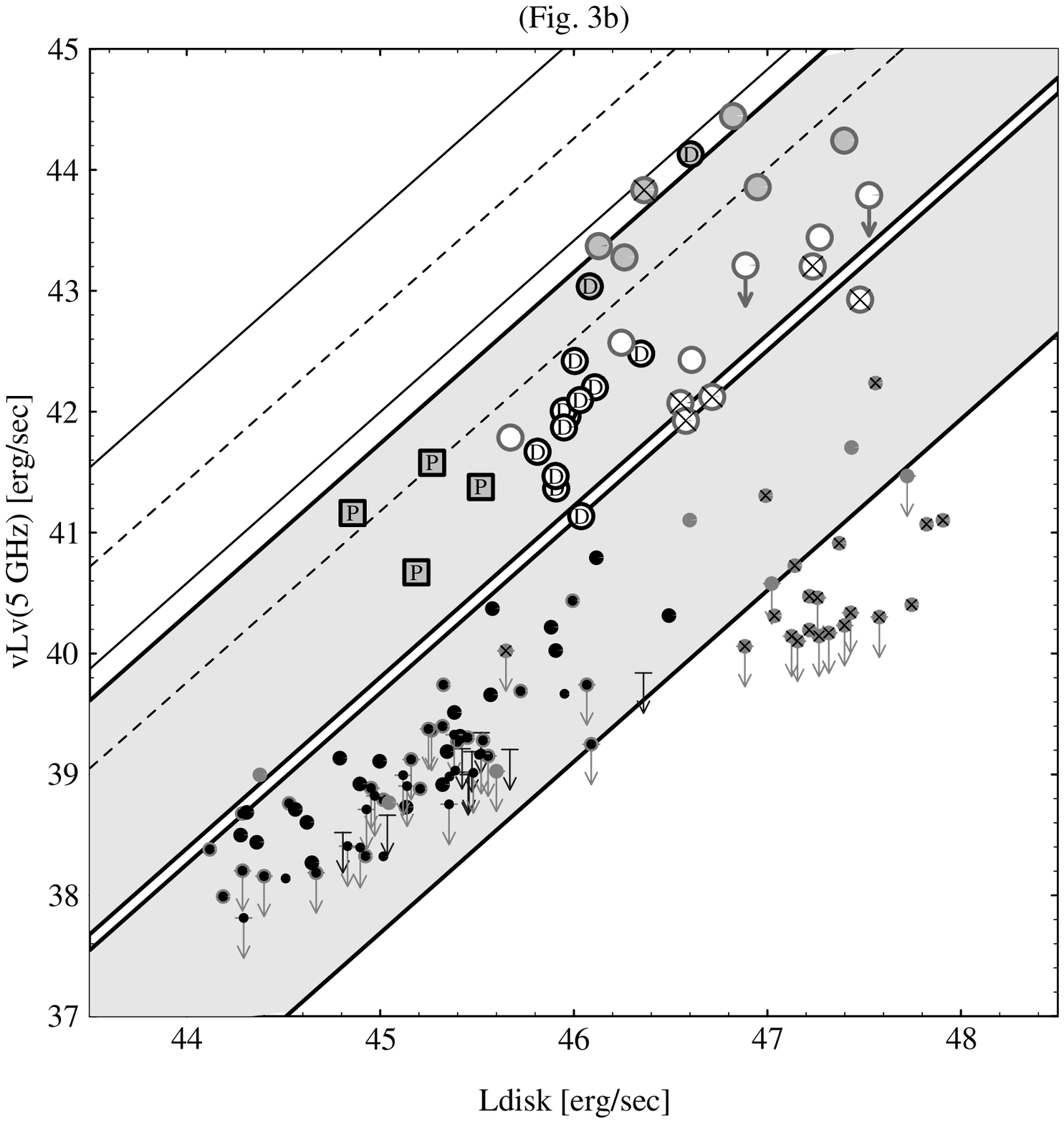}{8.5cm}{bbllx=2.7cm,bblly=5.3cm,bburx=20.0cm,bbury=22.2cm,clip=}}
\centerline{
\bildh{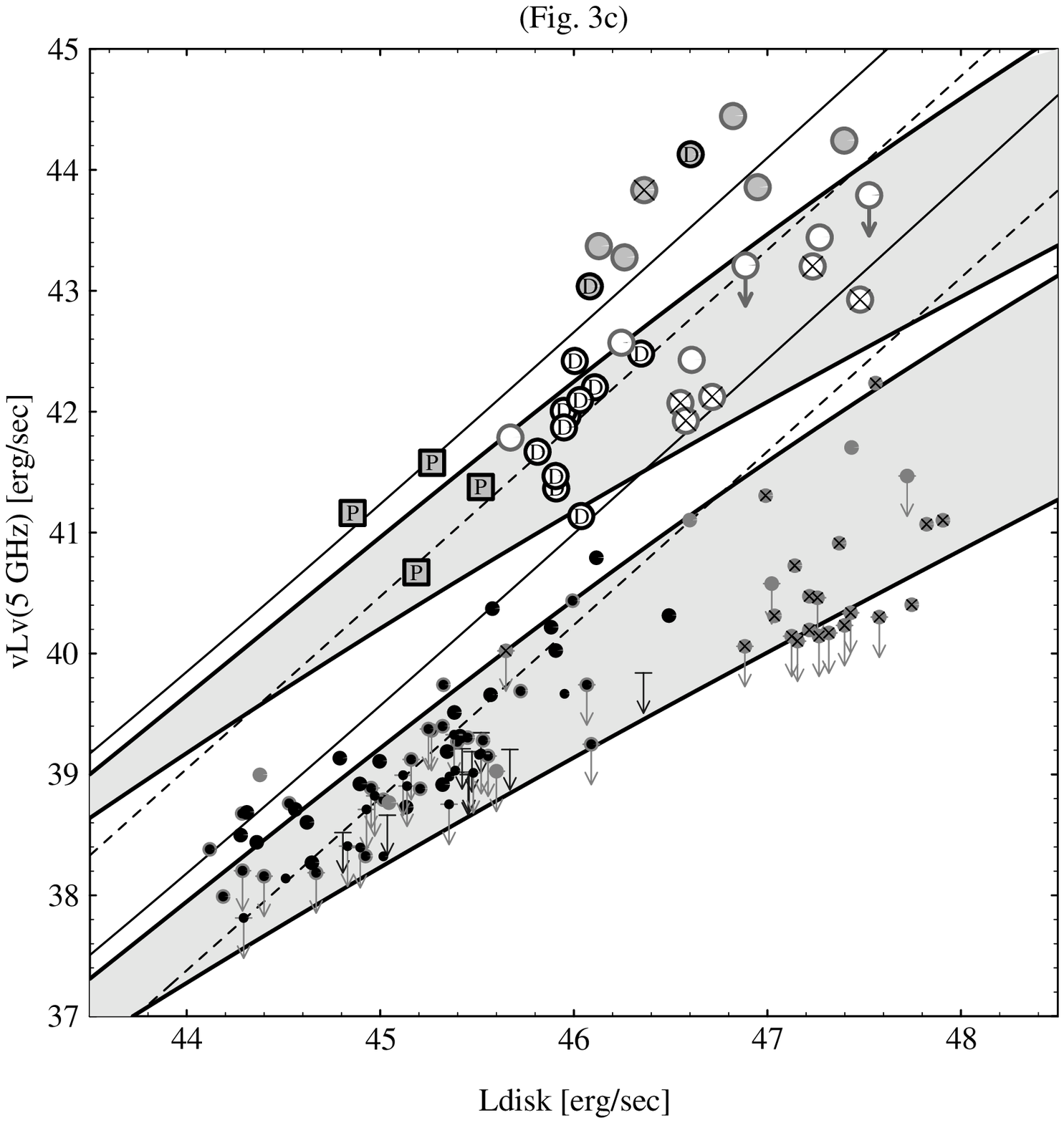}{8.5cm}{bbllx=2.7cm,bblly=5.3cm,bburx=20.0cm,bbury=22.2cm,clip=}
\bildh{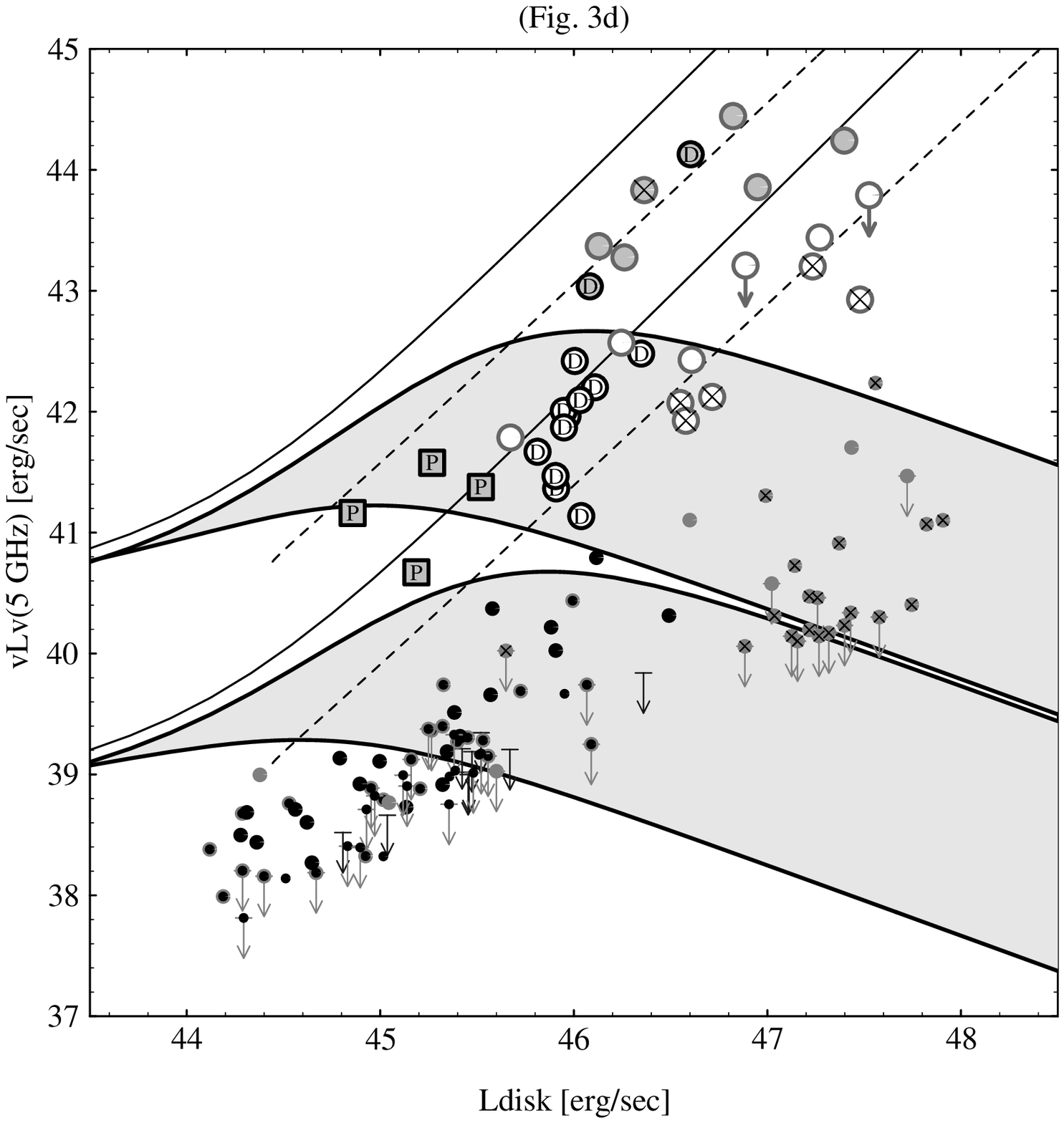}{8.5cm}{bbllx=2.7cm,bblly=5.3cm,bburx=20.0cm,bbury=22.2cm,clip=}}
 \caption[]{Different Models for the UV-radio correlation (Fig. 2) for
different Lorentz factors $\gamma_{\rm j,0}$ and velocity evolution
power index $\xi$. The jet power $q_{\rm j/l}$ is adjusted such that
the bulk of radio loud quasars at $L_{\rm disk}=10^{46}$ erg/sec are
always fit. Upper lane: `total equipartition' jet. Lower lane:
`protonic' jet. The widths of the grey lanes is given by different
inclination angle. Dashed line: Boosted sources with $i=1/\gamma_{\rm
j}$. Solid line: Maximum possible boosting ($i=0$). From top left to
bottom right we have: (a) Best fit: $\gamma_{\rm j}=6$, ($q_{\rm
j/l}=0.15$), $\xi=0.15$. (b) Lorentz factor too high: $\gamma_{\rm
j}=10$, ($q_{\rm j/l}=1$), $\xi=0$ (CDQ are too low). (c) Lorentz
factor too low: $q_{\rm j/l}=0.025$, $\gamma_{\rm j}=2.5$, $\xi=0.15$
(CDQ too high and distribution too narrow). (d) Velocity evolution too
strong: $\gamma_{\rm j}=6$, ($q_{\rm j/l}=0.15$), $\xi=0.5$.}
\end{figure*}

\subsection{Applying the parameters to radio weak jets}
As discussed in {\em Paper I} we proposed to test if the radio
emission from radio weak sources could be explained with emission from
jets similar to those observed in radio loud quasars. Based on the idea
that quasars are all powered by accretion disks around black holes,
it is hard to understand how the jet production in the accretion disk
very close to the black hole ($\ll1$ pc) should depend on
environmental effects (e.g. type of galaxy). We noted that a
difference in the efficiency of electron acceleration (e.g. production
of secondary pairs) could naturally account for the different radio
luminosities. The acceleration of these electrons must happen at a
larger scale ($\ga1$ pc) because of strong synchrotron losses
further in, and might well be triggered by environmental effects (e.g.
interaction between jet and ISM in a shear layer). But as radio weak
quasars, unlike their radio loud siblings, do not all populate the
extreme parts of the parameter space, observations are not yet able to decide
what is the real distinction between radio loud and radio weak.
In {\em Paper
I} we suggested at least another 3 alternative scenarios. For the
discussion in this Paper we will confine our attention only to the
case of a `protonic' jet ({\bf BP} model in {\em Paper I}).
Nevertheless, all other proposed models would yield similar results;
only the interpretation of the microscopic parameters would change.

The protonic jet is obtained by increasing the relativistic
proton/electron ratio $\mu_{\rm p/e}$ by a factor 100 which means we
reduce the energy stored in relativistic electrons by the same factor,
but leave everything else in the jet unchanged. The magnetic field
would still be in equipartition with relativistic particles, namely
the protons, and both still constitute a major fraction ($\sim50\%$)
of the total jet power (the other half is in kinetic motion). The
value for $\mu_{\rm p/e}$ is not ad-hoc but a natural configuration in
cases where all electrons are accelerated from their thermal
distribution ($x_{\rm e}\simeq1$) into a powerlaw distribution
starting at $\gamma_{\rm e,min}\approx1$.  From {\em Paper I} we
obtain as the core luminosity

\begin{eqnarray}\label{lnujmax3}
\nu L_\nu({\rm5 GHz})&=&1.0\cdot10^{41}\,{\rm erg/sec}\;
{\cal D}(L_{46},i_{\rm obs})^{2.17}\sin i_{\rm obs}^{0.17}\nonumber\\&&
\cdot\left({\gamma_{\rm e,min}x_{\rm e}}\right)^{0.83}
\left({6\over\gamma_{\rm j,0}}\right)^{1.8} q_{\rm j/l}^{1.42}
L_{46}^{1.42-\xi}.
\end{eqnarray}

There is no way to definitively prove this notion, but we could easily
falsify it. We just take the parameters derived for radio
loud jets, apply them to Eq.(\ref{lnujmax3}) and insert this in our
diagram without further fitting. The lower bands in Figure 3 represent
the protonic jet model where we only have changed the proton/electron
ratio by a factor 100 and the opening angles of the torus to $60\degr$
to be consistent with other findings as discussed before. The
agreement is surprisingly good and indeed we can easily match all
three populations (CDQ, LDQ, radio weak quasars) with a single model
where the difference is only due to relativistic boosting and
different relativistic electron populations.

We have to comment on one additional effect seen in this data. There
seems to be a slight curvature in the radio weak population towards
high powers. Unlike the case for the extended emission, there is no
reason to suppose that any kind of cosmological evolution in this
diagram would produce that effect.  The core luminosity depends only
on the accretion rate and not on the environment.  A change in the
accretion rate with $z$ should shift a quasar only parallel to the
distribution.  Strong evolution effects in this diagram would be a
serious argument to falsify our theoretical understanding of the radio
core.  However, the curvature is only minimal and could easily be
caused by technical or selection effects, especially as this curvature
is mainly seen in the high redshift PG sample where strong selection
effect are known to exist (Wampler \& Ponz 1985). Should selection
effects not be the main cause for this curvature one might argue that
this is indeed a sign of a velocity gradient index $\xi\neq0$, but
this is the only indication, and still $\xi=0$ gives a good
approximation to the data. One may test this in the future by
incorporating larger complete samples for radio loud and radio
weak quasars with high powers.

\subsection{RIQ: Boosted radio weak quasars?}
One inevitable consequence of our ``relativistic jets in all quasars''
hypothesis is that a fair fraction of all radio weak quasars are
boosted and appear as radio loud quasars. For the following discussion
we only refer to the PG sample with $z<0.5$ if not stated
otherwise.

The cores of the 11 steep spectrum LDQ appear as radio loud, these
quasars obviously form a separate subclass, which is physically
different from radio weak quasars already in the cores.  The rest of 5
objects, however, is a possible sample of beamed radio weak quasars,
because they show all characteristics one might expect from a boosted
source: strong compact core, high variability and flat spectrum. Two
of them (PG 1226+023 and PG 1302$-$102) show a lobe structure
themselves and therefore are likely to be boosted specimen of radio
loud quasars whereas the remaining 3 objects (PG 0007+106, PG
1309+355, PG 2209+184) apparently do not show any
extension and are candidates for boosted radio weak quasars. We use
these sources to define a subclass of radio quasars, the
radio-intermediate quasars (RIQ). To be classified as such we demand a
variable flat spectrum compact core without a large scale lobe
structure that in a UV-radio core diagram are found among, or on the
extrapolation of, the cores of LDQ. Two other candidates for this kind
of sources are PG 1351+640 and PG 1333+176 (see Notes).

We have now three independent means to check the idea that RIQ are
boosted radio weak quasars: the number of beamed sources, the offset
between boosted and normal quasars and the emission line
characteristics.

a) The ratio of the number of boosted to non-boosted objects reflects
the angle of the boosting cone which is $\arcsin 1/\gamma_{\rm j}$.
For a sample of $N$ objects with a boosting cone of $1/\gamma_{\rm j}$
and an obscuring dust torus with semi-opening angle of $\phi_{\rm
torus}$ we expect
\begin{equation}
n= N\left(1-\cos\arcsin\left(1/\gamma_{\rm j}\right)\right)/\cos\phi_{\rm
torus}
\end{equation}
strongly boosted objects. Here we have $N=73$, $n=3\pm1.7$ and assume
$\phi_{\rm torus}\simeq60\degr$ which corresponds to a $\gamma$-factor
for the jet of $\gamma_{\rm j}=5^{+2.7}_{-1}$. Inclusion of the fourth
source will reduce this value slightly.

b) The offset of the boosted from the non-boosted population is a
function of the Doppler factor as well. Alas, despite the major
popoulation being called `non-boosted' it is still affected by
boosting and we do not know exactly what the average inclination angle
of our sample really is, as we obviously have only upper limits in the
high inclination part of our diagram. We first fit a straight line
with {\em fixed slope} to both populations. The 3 boosting candidates
are a factor 153 more luminous in the radio. Using the Doppler factors
from our model, we find $\gamma$-factors in the range $\gamma_{\rm
j}=1.7$ to $\gamma_{\rm j}=4.8$ for the range of inclination angles
($60\degr-20\degr$) we attribute to the bulk of our parent population.
Here we assumed that all boosted jets show no inclination to the line
of sight; a more reasonable value for the inclination angle will lead
to slightly higher values of $\gamma_{\rm j}$. In Fig 3 the area
between the lower pair of dashed and full line represents the
expected region for boosted-radio weak quasars, for the parameter set
discussed before. Indeed, taking the uncertainties into account, the
RIQ are compatible with this model prediction. Of course, an
association with the extrapolation of the radio loud CDQ is not
completely excluded but is less likely.

c) BG92 found one significant emission line
indicator, which separated radio loud from radio quiet objects in so
far that radio loud objects show almost no evidence for Fe II emission
lines, whereas radio weak quasars do. Our sample of suspected boosted
radio weak quasars clearly shows enhanced level of Fe II emission, in
agreement with being radio weak, in marked contrast to the double lobe
sample which shows no significant FeII emission. In addition there are
some other differences between radio loud and radio quiet and taking
all together BG92 come to the conclusion that the sample
of flat spectrum sources has emission properties like radio quiet
quasars but unlike radio loud quasars.

The hypothesis of boosted radio weak quasars passed all three tests
reasonably well. Both estimates for the Lorentz factor are in
agreement with each other and our previous estimate and support the
hypothesis of having relativistic jets in radio weak quasars.

\subsection{FIR emission}
Another consequence of having jet emission in radio weak quasars is
the possibility to see this emission at higher frequencies. In {\em
Paper I} we showed that we can expect non-thermal emission in the rest
frame up into the submm regime. Unfortunately the expected flux is on
the mJy level which is difficult to observe with current detectors.
Chini et al.  (1989) observed a subsample of the PG quasars at 230
GHz, showing that the IR emission at higher frequencies in quasars is
of thermal origin.  They detected only 5 quasars at 230 GHz and
concluded from the spectrum that the emission up to this frequency is
non-thermal while beyond 230 GHz thermal dust emission takes over. We give a
table of spectral indices $\alpha$ between these fluxes and the 5 GHz
VLA core and total fluxes (except for 2041$-$109 where we take the flux
referenced in Chini et al. 1989) of the 5 objects being in our
extended sample. We define the spectral index to be positive for
inverted spectra ($F_\nu\propto\nu^{\alpha}$).

\begin{table}
 \caption[]{High frequency spectral indices for radio weak quasars
detected at 1.3 mm: $\alpha_{\rm core}$ is the spectral index between
the coreflux at 6 cm and the total flux at 1.3 mm while $\alpha_{\rm
total}$ is the spectral index between the total flux at 6 cm and 1.3
mm.}
\begin{center}
\renewcommand{\arraystretch}{1.3}
\begin{tabular}{|llll|}
\hline
Name&$F_{230 {\rm GHz}}[mJy]$&$\alpha_{\rm core}$&$\alpha_{\rm total}$\\
\hline
\hline
0007+106&$3.9\pm1$&$0.51^{+0.06}_{-0.08}$&$\hphantom{-}0.48^{+0.06}_{-0.08}$\\

1411+44&$3.9\pm1$&$0.51^{+0.06}_{-0.08}$&$\hphantom{-}0.48^{+0.06}_{-0.08}$\\
2041--10&$4.0\pm1$&$-$&$-0.08^{+0.05}_{-0.08}$\\
2130+09&$2.1\pm0.7$&$0.23^{+0.08}_{-0.1}$&$\hphantom{-}0.01^{+0.08}_{-0.1}$\\
2214+13&$3.5\pm0.6$&$-$&$\hphantom{-}0.7^{+0.04}_{-0.05}$\\
\hline
\end{tabular}
\end{center}
\end{table}

All spectral indices are flat or inverted as expected if there indeed
is a strong flat (inverted) spectrum core. In our picture the
non-thermal emission at this frequency should come from a very small
region of the order $10^{17}$ cm away from the disk and therefore we
are allowed to peek into the region where the jet establishes itself
and may still be in a state of accelerating its inital electron
population. Interestingly the Galactic Center source Sgr A* shows a
similar inverted spectrum in the mm to submm range where we explained
this qualitatively as being radiation coming from the jet nozzle (FMB,
Falcke \& Biermann 1994a, Falcke 1994b).

\section{Unification with FR\,II galaxies}
\subsection{Estimating the hidden disk luminosity}
An often used assumption is that FR\,II radio galaxies and quasars are
basically the same objects but seen at different inclination angles
(Barthel 1989).  Seen edge on, the obscuring dust torus will prevent
us from observing UV-bump and broad line region and we cannot classify
the radio source as a quasar. In fact, if an obscuring dust torus is
indeed present, and the widths of the lanes in Fig. 3 do support such
a concept, the separation between quasars and radio galaxies is a
necessary consequence.

In the following we will assume that the unification scheme is correct
and that FR\,II galaxies and radio loud quasars are identical except
for different inclination angles. This implies the presence of an
accretion disk in FR\,II radio galaxies like in quasars and we would
expect that inserted in our radio/$L_{\rm disk}$ diagram the cores of
FR\,II radio galaxies should occupy a region slightly below the radio
loud quasars -- because of less boosting -- while the total (lobe)
emission radio galaxies and quasars should occupy exactly the same
region. For testing this we will make use of the sample of radio
selected FR\,II galaxies used by RS91.

A big problem of the sources in this sample is that per definition it
is almost impossible to measure their disk luminosity.  Rawlings et
al. (1989) tried to use the O{\sc III} narrow line as an indicator but
found only weak correlations. Later RS91 used a combination of
different narrow lines to obtain a better indicator for the central
luminosity and indeed found a good correlation between the luminosity
of these lines and the power stored in the radio lobes divided by the
age of the lobe -- the total jet power.  Unfortunately, they did not
tabulate these narrow line luminosities, and anyway it would be
difficult to calibrate such an indicator to an absolute value of the
disk luminosity as we did for our sample.

Hence, we are not able to construct a real radio/$L_{\rm disk}$
diagram for FR\,II radio galaxies and prove any optical/radio
correlation. Nevertheless, we can use the knowledge gained in this
paper to perform an important consistency check of our results.  The
idea, we have followed throughout this paper, is that the average
ratio between jet power and disk luminosity is constant within our
sample. Given that the unification scheme and the value $Q_{\rm
jet}/L_{\rm disk}$ we have derived is correct we can now turn the
argument around and estimate the hidden disk luminosity of FR\,II
radio galaxies from the jet power. For this we take the jet power
$Q_{\rm jet,RS91}$ derived by RS91 and correct their minimum energy
estimates by a factor $f_{\rm RB93}\sim3$ for the presence of
relativistic protons.  Rachen \& Biermann (1993, RB93) and Rachen et
al. (1993) investigated the extragalactic ultra-high energy tail of
the cosmic ray population, which is most likely to stem from radio
lobes, and found that the cosmic ray ratio in hotspots of FR\,II
galaxies should be of the order 20 which yields a factor 3 increase
for the derived total jet power. Thus we obtain as a conversion formula

\begin{equation}\label{ldiskfrii}
L_{\rm disk}=10 \; Q_{\rm jet,RS91} \left({f_{\rm
RB93}\over3}\right)\left({q_{\rm j/l}\over 0.15}\right)^{-1}.
\end{equation}
where a factor $2 q_{\rm j/l}$ accounts for the two-sidedness of the
jets.

\subsection{Core emission}

Now we can use the hidden disk luminosity Eq.(\ref{ldiskfrii}),
proceed as before, and plot radio core luminosities versus disk
luminosity. We used the core fluxes listed in Rawlings et al. (1989)
and total fluxes from the NED database and converted them into the
rest frame just the way we did for quasars.
\begin{figure}
\centerline{\bildh{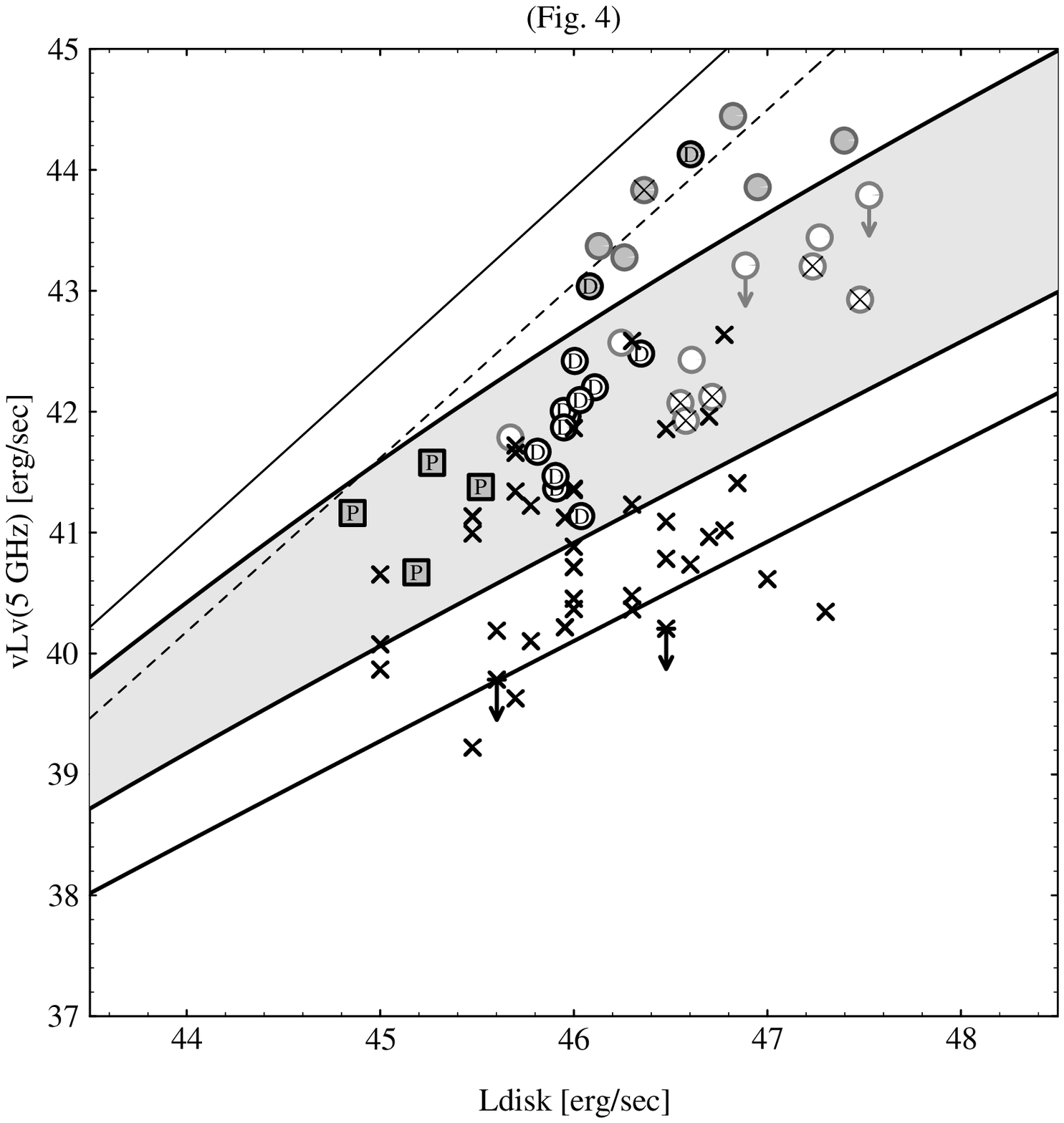}{8.5cm}{bbllx=2.7cm,bblly=5.3cm,bburx=20.0cm,bbury=22.2cm,clip=}}
\caption[]{Monochromatic luminosity of the compact radio core versus the
disk luminosity (see text) for radio loud quasars with (circles and
boxes -- see Fig. 2) and FR\,II galaxies with $z<0.5$ (crosses) . The
`disk luminosity' for FR\,II galaxies was obtained by applying a general
conversion factor $L_{\rm disk}=10 Q_{\rm jet}$ to their lobe energies
estimated by RS91. The shaded band gives our radio loud model for
inclination angles of $15\degr$ to $45\degr$. The lower grey line is
the extension of the band for $90\degr$ inclination.}
\end{figure}

With the conversion factor $q_{\rm j/l}=0.15$ and $f_{\rm
RB93}=3$ we reproduced our diagram in Fig. 4 for the radio loud sources
only, where small crosses represent the FR\,II cores. The theoretical
`best fit' radio loud model from Fig. 3 is redrawn and the grey line
below gives the widening of the band for including up to $90\degr$
(edge-on) orientation.

The expected coincidence between FR\,II radio galaxies and quasars is
indeed present.  The overlap of these two populations was obtained
without further adjusting previously constrained parameters and
therefore is an independent test for several conclusions drawn before,
e.g. the high jet power $2 Q_{\rm jet}/L_{\rm disk}\sim0.3$. This is
also a check for the consistency of the unification theory, and we
conclude from the fact that FR\,II galaxies occupy the anticipated
region that the implied scaling of the jet cores with disk luminosity
seems to be correct in quasars, as well as in FR\,II galaxies and that
double lobed quasars and FR\,II galaxies have similar jet properties.
A major parameter to influence the core brightness is the inclination
angle, and as expected the FR\,II cores cluster at the lower end of
the distribution, such that approximately $2/3$ of the FR\,II cores
are close to or below the $45\degr$ line of the model extending well
down to the $90\degr$ line.

This effect is highlighted by the core to lobe ratio at 5 GHz
(restframe) in both populations for the quasars where the median values are
\begin{equation}\label{core/lobe}
\left<{L_{\rm core,quasars}\over L_{\rm lobe,quasars}}\right>=15\%\;\;\;\;\;
\left<{L_{\rm core,FRII}\over L_{\rm lobe,FRII}}\right>=0.8\%\;.
\end{equation}
If we exclude the two CDQ one would get $8\%$ for the quasars, still
the quasars have substantially brighter cores compared to FR\,II
galaxies.  This originally was one of the reasons to propose the
quasar/FRII unification scheme. Only one of the radio cores (3C341)
falls far below the theoretically expected range and one should
re-examine the determination of the jet power or look for other
pecularities. On the other hand this offset might only indicate the
larger uncertainty in determining the hidden disk luminosity.

\subsection{Extended emission}
Does the coincidence of FR\,II and quasar fluxes also hold if one
compares the extended fluxes? Figure 5 shows the total flux of all
sources plotted versus the disk luminosity. FR\,II sources are again
incorporated using the conversion parameter $2 q_{\rm j/l}=0.3$ as
before. We used the total luminosity as it is in most cases dominated by the
lobes and it is by far the most reliable datum we have. As the lobe
emission is assumed to be unaffected by boosting we expect that the
total flux of LDQ and FR\,II show exactly the same distribution. The
fact that this is indeed the case is another independent check of the
consistency of unification and of our parameter $q_{\rm j/l}$. The
distribution scales with the disk luminosity and is very narrow; it
scatters only by a factor 3 in both directions -- much less than for
the core fluxes where different inclination angles broaden the
distribution. CDQ are still at the upper end of the distribution,
however, they are no longer clearly separated.

We also plotted as a full line the lobe luminosity $2\cdot L^*_{\rm
lobe}$ derived in {\em Paper I} from extrapolating the physical
situation in the core out to the lobes; the dashed lines are each
offset by 0.5 in the log to indicate the scatter. Although the
agreement is fairly good one should be careful with the interpretation
as the derivation is very simplified (e.g. the dissipation of kinetic
energy is still unclear) and serves only as an order of magnitude
check.

\begin{figure}
\centerline{\bildh{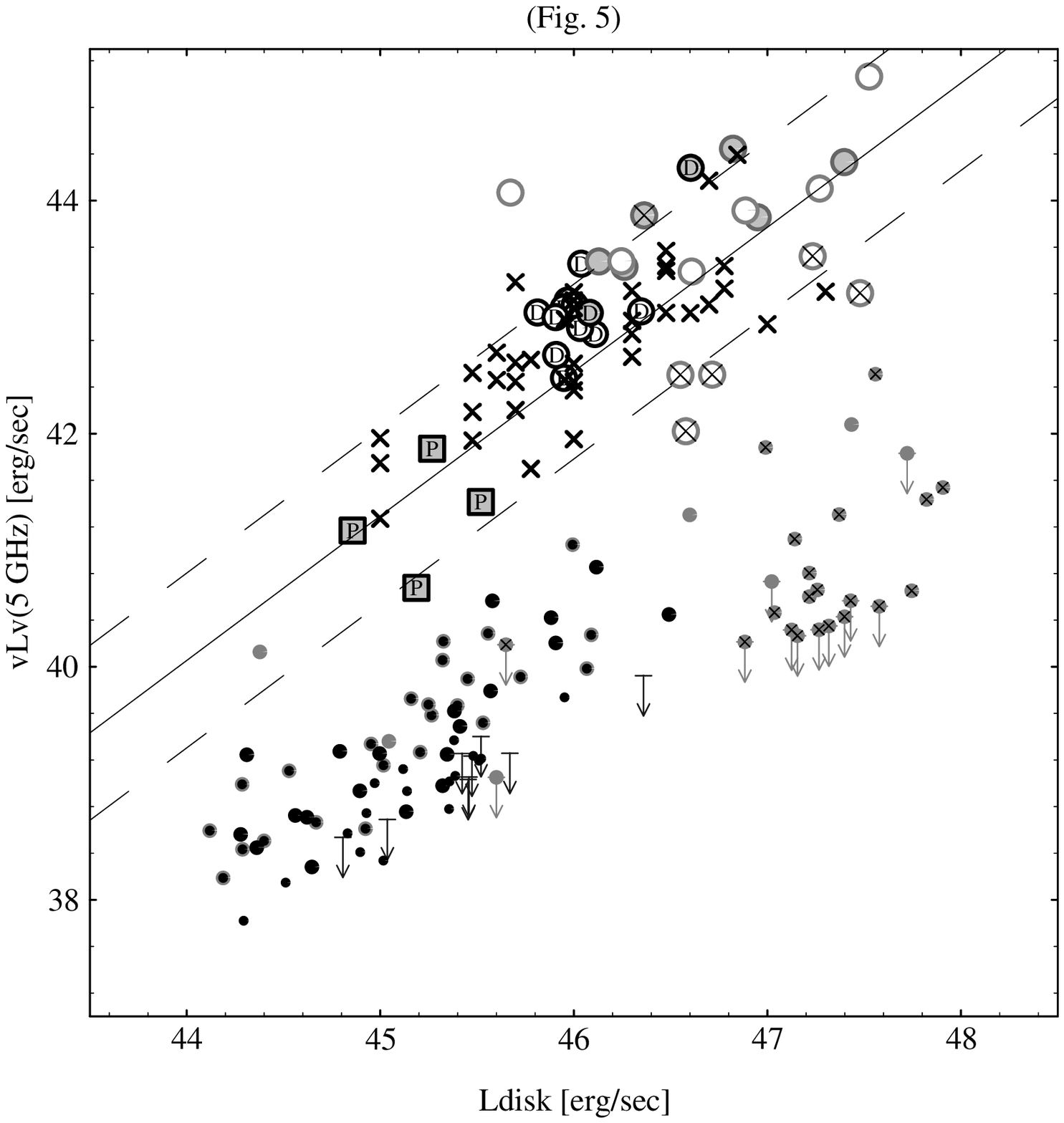}{8.5cm}{bbllx=2.7cm,bblly=5.3cm,bburx=20.0cm,bbury=22.2cm,clip=} }
 \caption[]{Monochromatic total luminosity of quasars (boxes, circles
and points -- see Fig. 2) and FR\,II galaxies (crosses).  The `disk
luminosity' for FR\,II galaxies was obtained by applying a general
conversion factor basing on our analysis to the jet powers (RS91). The
radio loud quasar sticking out from the general distribution is the
CSS source 3C48.}
\end{figure}

In our sample consisting of sources with $z<0.5$ we do not expect
strong cosmological evolution, for sources with higher redshift it
appears as if indeed there is a tendency for PG quasars to have lower
radio flux, perhaps caused by cosmological evolution. Lobes of quasars
at higher redshift may be relatively fainter although equally bright
in absolute terms, but statistical statements seem to be problematic
with this small number of sources and it is suspicious that radio loud
and radio weak high redshift PG quasars seem both to be shifted to
relatively lower radio emission whereas the other soures do not show
this trend.

\subsection{Extended emission in radio weak quasars}
For the radio weak population there also is a correlation between
total radio luminosity at 5 GHz and disk luminosity, but the scatter of
the distribution is larger than in the radio loud quasars and slightly larger
than in the cores alone. In these sources we obviously have a mixture
of many contributions. A popular idea is that the diffuse emission is
caused by starbursts or starburst-driven superwinds (Baum et al.
1993), but in this case starburst and AGN activity must be strongly
linked by the large scale accretion flow in the galaxy to provide this
correlation.  The alternative explanation than is that a large
fraction of the extended emission is caused by diffuse emission from
the terminating or disrupting jet flow as seen in many Seyfert
galaxies (e.g. NGC1275, see B\"ohringer et al. 1993 and refs.). This
would require the jet to terminate at the kiloparsec scale, which is
difficult to explain with the assumption of equal powers for jets in
radio loud and radio weak quasars. One would have to favour a
dichotomy in the jet production itself with a lower $Q_{\rm jet}/L_{\rm
disk}$ in radio weak quasars and not a dichotomy in the electron
acceleration. On the other hand radio weak quasars are mostly spirals
having a completely different density profile along the rotation axis
as ellipticals where radio loud quasars are found, and one might
speculate that jets in spirals either terminate almost invisibly far
out or very close to the center. Interestingly half of the
detected radio weak quasars show no extended emission at all.

\section{Summary}
We used an optically selected quasar sample to test our idea that jet
and disk are symbiotic features which are equally important for the
overall energy budget of the AGN and are present in radio loud and
radio weak quasars. From UV-bump fits we derived empirical relations
between disk luminosity and other luminosity indicators (optical
continuum, $M_{\rm b}$, O{\sc III}, He {\sc II} and ${\rm
H}_{\beta}$). We then plotted the average disk luminosity of each
quasar versus the radio luminosity at 5 GHz of its compact core and
find good correlations. This suggests that {\em compact core radio
emission and the UV bump have a common energy source in all quasars}.
Using the disk luminosity one has an absolute reference scale to
compare objects according to their accretion power. In our UV-radio
diagram the quasars split into 4 groups separated by their radio core
brightness relatively to the disk luminosity: core dominated quasars
(CDQ), lobe dominated quasars (LDQ), radio-intermediate quasars (RIQ)
and radio weak quasars.
\begin{itemize}
\item{For a given disk luminosity the cores of radio loud quasars are brighter
than those of radio weak quasars, hence the difference between radio loud and
radio weak quasars is established already in the inner parsecs.}
\item{For a given disk luminosity CDQ have the brightest cores due to
relativistic boosting.}
\item{AGN must have a disk luminosity $\ga10^{46}$ erg/sec to be able to become
a radio loud quasar and produce a FR\,II type jet.}
 \item{The radio core emission of radio weak quasars scales with the
central disk luminosity as well.}

 \item{A subsample of flat spectrum variable
point sources (RIQ) is substantially fainter than CDQ but almost as
bright as the cores of LDQ in a disk luminosity regime $<10^{46}$
erg/sec, we interpret these sources as boosted radio weak quasars}

 \item{The one CSS (3C48) quasar in our sample has a core luminosity
comparable to other quasars of the same disk luminosity but a
tremendously enhanced total radio flux -- the jet is undisturbed in
the core but disrupted by environmental effects further out}
\end{itemize}
Together with our previously developed model for the jet core
emission, where we added energy and mass conservation of the jet-disk
system to the standard emission model, we now can strongly constrain
the parameter range for jet models.

\begin{itemize}
 \item{Radio loud cores are the most efficient case of a jet and, given
the constraints from the disk luminosity, can only be explained if the
energy in magnetic field and relativistic particles represent a large
fraction ($\sim50\%$) of the total jet energy.}
\item{This high power
of the relativistic particle distribution in radio loud cores can not
be explained by accelerating thermal plasma electrons smoothly into a
powerlaw distribution but requires a low-energy cut-off of the
relativistic electrons around 50 MeV or the presence of a large
population of pairs or both.}
\item{The power of the jet approaches  1/3 of the disk lumi\-nosity,
therefore the jet is energetically important for the disk structure and
evolution.}
\item{The bulk Lorentz factor of the jets are between 3 and 10 and
can not be much higher because Doppler boosting would reduce the
radio luminosity of the bulk of the poulation too much, and the width
of the distribution would be strongly increased by the random
orientation of inclination angles.}
\item{For the same reason the dependence of the bulk Lorentz factors
on the disk luminosity must not be very strong ($\gamma_{\rm j}\propto
L_{\rm disk}^\xi$ and $\left|\xi\right|\ll0.5$).}
\item{The distribution of cores of radio weak quasars can be
explained with the same type of relativistic jets as in radio loud
quasars but either the ratio $Q_{\rm jet}/L_{\rm disk}$ is lower or
radio weak quasars simply do not reach the extreme efficiency in
accelerating electrons.}
\item{The properties of CDQ and RIQ are consistent with being the
boosted counterparts of radio loud and radio weak quasars
respectively.}
\end{itemize}

We conclude that the difference between radio loud and radio weak
quasars need not necessarily arise from different kinds of jets and
disks but may simply be due to the presence of a highly energetic
electron population in the jets of radio loud quasars. Given the
limits from mass and energy conservation in the accretion flow and
assuming that electrons are accelerated from the thermal regime into a
powerlaw distribution it is impossible to get more radio emission than
what is seen in radio weak quasars. To produce radio loud cores an
additional process is required that either shifts the whole
relativistic electron distribution towards higher energies or creates
lots of additional particles, i.e.  pairs.  In {\em Paper I} we
suggested that hadronic cascades initiated by high energy proton $pp$
or $p\gamma$ collisions might be such a process that could be
additionally switched on in radio loud quasars (e.g. due to
interactions in the shear layer between jet and ISM). Interestingly,
hadronic cascades could explain the high energy gamma-ray emission
seen in these objects as well (Mannheim 1993). Secondary pair
production was also already successfully applied to model the radio
emission of the old nova GK Per (Biermann, Strom, Falcke 1994).

With our estimate of the ratio between jet power and disk luminosity,
we are also able to estimate the hidden disk luminosity in FR\,II
radio galaxies. By virtue of this method we can insert FR\,II sources
in our UV-radio correlation and show that LDQ and FR\,II cores as well
as LDQ lobes and FR\,II lobes coincide in this diagram. This
demonstrates consistency of our parameter set and of the FR\,II/quasar
unification scheme. In this context we point out that the lack of
FR\,I quasars in our sample -- especially in the low power regime
below $10^{46}$ erg/sec -- shows that the difference between FR\,I and
FR\,II can not simply be due to a decrease in disk or jet power alone.
As noted already by MRS there appears to be a critical power of
$L_{\rm disk}\approx10^{46}$ erg/sec below which only extremely few --
if any -- radio loud quasars with FR\,II morphology exist. This might
be an important clue for our understanding of the FR\,I/FR\,II
dichotomy (see Falcke, Gopal-Krishna, Biermann 1994).

\begin{acknowledgements}
HF is supported by the DFG (Bi 191/9). We thank M. Rieke and G. Rieke
for valuable suggestions and comments. Continuous discussions with J.
Rachen, B. Nath and M. Niemeyer helped to clarify some subtle points
in our arguments. We also profited from numerous discussions on
`unified theory' and radio galaxies with Gopal-Krishna. This research
has made use of the NASA/IPAC extragalactic database (NED) which is
operated by the Jet Propulsion Laboratory, Caltech, under contract
with the National Aeronautics and Space Administration.
\end{acknowledgements}

\appendix
\section{Appendix: Notes on individual sources}
{\it PG 0007+106 (III Zw 2):} A very unusual quasar which attracted a
lot of interest in the past (Kaastra \& Korte 1988 and refs. therein).
The radio flux density varies more than a factor 30 and shows flare
like events on a timescale of one year.  Variability is found at all
other wavelengths as well from infrared to x-rays.  Past VLBI
experiments found some 80\% of the total flux in a compact core and a
jet like extension at PA $148^{\circ}$ (no maps are presented). Unger
et al.~found a weak component 30 kpc away from the core with the VLA
at 1.5 GHz not seen at 5 GHz. The core flux density has varied from
425 mJy to 40 mJy at 8.4 GHz between June 1991 and Oct.~1992 (Patnaik,
A., p.c.). Large and fast changes of the polarization angles at 4.8
GHz and 8 GHz, outbursts on a month timescale and variability $<10$
days at 37 GHz (no polarization measurements) have been found (Aller
et al.  1985, Ter\"asranta et al.  1992).  These are typical
signs for boosted jet emission, however, III Zw 2 seems to fall on the
extrapolation of the lobe dominated quasars and not of the CDQ. We
consider this source as the archetype of the RIQ.

{\it PG 0044+030:} This source has a relatively low total flux and
appears diffuse elongated on a recent VLA map at 20 cm while at 6 cm
only the core prevails (Price et al. 1993).

{\it PG 0157+001:} We regard the core flux given by KSSSG (5.9 mJy) of
this compact source as an upper limit for a flat spectrum core.  The
spectrum by Barvainis \& Antonucci (1989) of the VLA peak flux does
show a very steep spectral index ($\alpha\sim-1$) and there is no
indication of a flat component above 1 mJy. In adition the KSSSG core
flux would make 0157+001 also a possible candidate for being a RIQ
which we think is not justified in light of the spectral form. On the
other hand 0157+001 has a host galaxy which is almost face on (Hughes et
al. 1993) and hence if the jet axis is related to the rotation axis of
the galaxy there is a high probability for getting enhanced emission
due to boosting.

{\it 0232-042:} A triple radio source at VLBI scales with a strong
bent. The core flux is from Saikia et al. (1984).

{\it 0237-233:} This is a GPS with two components (lobes?) at the VLBI
scale. The spectrum continues to be steep up to 31 GHz and we can only
estimate an upper limit for a flat component of $f_{\rm c}\approx0.1$
which seems to be consistent with the overall picture.

{\it 0134+329, 3C 48:} is a famous CSS. High resolution maps (Wilkinson
et al. 1991) show a compact core and a hotspot plus a huge amount of
diffuse emission making this source outstandingly bright in total
flux.  Due to an unusually low core/lobe ratio of $0.007$ the core
flux is consistent with other radio loud objects. This strengthens the
idea that 3C 48 is powered by an ordinary radio jet where disruption
by environmental effects close to the nucleus causes an enhancement of
the (lobe) radio emission.

{\it 0414-060, 3C 110: } This source is unresolved (Ekers 1969) and the
mean value of .11 for the lobe to core ratio we assigned to this
source may be inappropriate (see 3C48) but the spectrum is steep with
a slight flattening at higher frequencies which might indicate the
presence of a weak flat component.

{\it 0710+457, Mkn 376:} We took 1.6 mJy as VLA core flux at 5 GHz from
Neff \& Hutchings (1992). The source is compact ($<0.8\arcsec$) and if
the total flux of 21 mJy is correct than Mkn 376 has relatively bright
extended emission.

{\it 0955+326, 3C232:} Akuyer et al. (1994) show a map of a compact
$\sim1$\arcsec source and quote a flattening of the spectrum to
$\alpha=-0.2$ between 1.4 GHz and 408 MHz therefore it might be a
compact steep spectrum source (CSS) or even a GHz peaked source (GPS)
and the assumed value for the core lobe ratio is very uncertain
although the source fits into the general trend. The total emission
however is at the upper end of the distribution

{\it PG 1008+133:} UV-bump fit and $M_{\rm b}$ yield disk luminosities
which differ by more than a factor 20 so that the average disk
luminosity is completely uncertain. We excluded this source from our
fit to the $L_{\rm disk}-M_{\rm b}$ correlation.

{\it PG 1211+143:} This source was obviously misidentified in KSSSG
and we take the fluxes given by MRS.

{\it PG 1206+459:} UV-bump fit and $M_{\rm b}$ disk estimates differ
by a factor 100 (same as PG 1008+133).

{\it PG 1309+335:} MRS report a total flux density of 54 mJy
concentrated in a single core component. The flux was $\sim35-50\%$
brighter than in previous surveys (Gregory \& Condon 1991: 40 mJy,
Becker et al. 1991: 35 mJy).  The spectrum is flat and it is a point
source at the VLA. In our diagram it occupies the region around III Zw
2; we consider it a RIQ as well.

{\it PG 1333+176} has a core/lobe ratio of 1.2 in KSSSG indicating
variability and very weak extended emission. Nevertheless this source
is at the lower end of the radio loud sources and one might speculate
that this is another example of a RIQ quasar. The total emission is
at the extreme lower end for radio loud sources.

{\it PG 1351+640:} This is a variable point source. The flux dropped
by a factor of 6 within 4 years (Barvainis \& Antonucci 1989). In Fig.
1 it clearly stands out from the normal population but does not quite
reach the RIQ. One is tempted to call this source an intermediate RIQ
source. The fact that the light curve showed a steady decline over 4
years and not a wiggling around an average value makes it difficult to
classify the source at present beacuse our averaging procedure gives a
strongly biased result. Using the data of KSSSG only would lower the
radio flux substantially although it would still be well above the
radio weak population. Like 1333+176 one will need more data to come
to an unambigious conclusion. Interestingly McLeod \& Rieke (1994)
find that this is one out of three sources where the luminosity
profile is better fitted by an elliptical than by a disk galaxy.

{\it PG 1407+265:} was classified as radio loud by
Sanders et al.  (1989) which is in clear contradiction to its position
in Fig. 1 and we labeled it radio weak especially as the core/lobe
ratio ($\sim0.4$) is fairly high.

{\it 1501+106:} This source had originally a fitted disk luminosity
which was a factor 10 higher than what we estimated from the other
luminosity indicators because it was the only object where SM89 made
an extreme (high $\dot M$) fit to try to hit the strong soft x-ray
excess. We now redid this and went only for a good fit to the
optical/UV data which lowered the disk luminosity estimate
substantially.

{\it PG 1634+706:} Barvainis \& Antonucci (1989) find a slightly inverted
spectrum for the VLA peak flux; thus at least here the assumption of a
flat spectrum core seems to be justified.

{\it PG 1700+518:} A broad absorption line (BAL) quasar like PG
1411+442, was labeled a RIQ by MRS because of its high total flux
which is caused by two compact knots separated by 0.9\arcsec where the
optical nucleus is close to the brighter radio peak. Considering its
core flux this source is not outstanding at all and also lacks the
other characteristic features of a RIQ as we define it (pointlike
appearance and variability). It appears more likely that like in CSS
and GPS environmental effects cause the enhancement of the total radio
flux by disrupting a weak radio jet.

{\it E1821+643:} It is probably a giant elliptical galaxy in the
center of a rich cluster with a very luminous AGN. Radio core and jet
structure are seen in the center but the total radio flux is very low
(Hills et al. 1992). In our diagram it is placed on the upper end of
the radio weak distribution supporting the impression that this is a
radio weak jet in an environment usually producing radio loud quasars.

{\it 2041-109,Mkn 509:} VLA data by Barvainis \& Antonucci (1989)
indicate a flattening of the spectrum beyond 5GHz so that there is
indeed a flat component. We took 1.8 mJy as core flux at 5GHz from
Neff \& Hutchings (1992).

{\it PG 2209+184:} VLA total fluxes for this source (z=0.07) at 5 GHz
are reported to be between 120 mJy (MRS) and 290 mJy (KSSSG, claimed
to be in error by Miller et al.)  with a VLA core flux $>95\%$ of the
total flux and a low brightness knot ($S_{4.8{\rm GHz
}}\simeq3$mJy/beam). The spectrum is inverted between 5 GHz and 1.4
GHz ($\alpha=-0.24$). Only recently Machalski \& Magdziarz (1993)
explicitly identified this source as variable with total fluxes at
5GHz between 116 mJy and 326 mJy. Like III Zw 2 we classify it as a
RIQ.

\end{document}